\documentclass{aastex}

\shorttitle{Mid-Infrared Imaging of ULIRGs}
\shortauthors{Soifer  et al.}

\begin{document}

\title{High Resolution Mid-Infrared Imaging of Ultraluminous Infrared Galaxies 
\altaffilmark{1}
}

\author{B. T. Soifer\altaffilmark{2}, G. Neugebauer, K. Matthews, E.
Egami} 
\affil{Palomar Observatory, California Institute of 
Technology, 320-47, Pasadena, CA 91125}
\email { bts@mop.caltech.edu, gxn@mop.caltech.edu, kym@mop.caltech.edu, 
egami@mop.caltech.edu}
\author{E. E. Becklin, A. J. Weinberger} 
\affil{Department of Physics and Astronomy, University of California Los
Angeles, 156205 Los~Angeles, CA 90095}
\email {becklin@astro.ucla.edu,alycia@astro.ucla.edu}
\author {M. Ressler, M. W. Werner}
\affil {Jet Propulsion Lab, 169-506, 4800 Oak Grove Dr., Pasadena, CA 91109}
\email {ressler@cougar.jpl.nasa.gov,mww@ipac.caltech.edu}
\author {A. S. Evans\altaffilmark{3} , N.Z. Scoville}
\affil {Divison of Physics, Mathematics and Astronomy, California Institute of 
Technology, 105-24, Pasadena, CA 91125}
\email {ase@astro.caltech.edu, nzs@astro.caltech.edu}
\author {J.A. Surace}
\affil {SIRTF Science Center, California Institute of Technology, 314-6, 
Pasadena, CA 91125}
\email {jason@ipac.caltech.edu}
\and
\author { J.J. Condon}
\affil {National Radio Astronomy Observatory, 520 Edgemont Road,
Charlottesville, VA 22903}
\email { jcondon@nrao.edu}

\altaffiltext{1} {Based in part on observations obtained at the
W. M. Keck Observatory  which is operated as a scientific partnership
among the  California Institute of Technology, the  University of
California and the National Aeronautics and Space Administration.}

\altaffiltext{2} { also at SIRTF Science Center, California Institute of 
Technology, 314-6, Pasadena, CA 91125}
\altaffiltext{3} { current address: Department of Physics and Astronomy,
SUNY Stony Brook, Stony Brook NY 11794-3800}
\vfill\eject

\begin{abstract}
Observations  of ultraluminous infrared galaxies (ULIRGs) with an 
achieved resolution approaching 
the diffraction limit in the mid-infrared from 8 - 25 $\mu$m using the
Keck Telescopes are reported. We find extremely compact structures,
with spatial scales of $< 0.3''$ (diameter) in six of the  seven  ULIRGs
observed. These compact sources emit between 30\% and 100\% of the
mid-infrared energy from these galaxies.   

We have utilized the compact mid-infrared structures as a diagnostic
of whether an AGN or a compact (100 -- 300 pc) starburst is the
primary power source in  these ULIRGs. In Markarian 231, the upper
limit on the diameter of the 12.5\,$\mu$m source , 0.13$''$, shows
that the size of the infrared source must increase with increasing
wavelength, consistent with  AGN models. In IRAS~05189-2524 and
IRAS~08572+3915 there is strong evidence that the source size
increases with increasing wavelength.  This suggests heating by a
central source rather than an extended luminosity source, consistent
with the optical classification as an AGN.

The compact mid-infrared sources seen in the other galaxies cannot be
used to distinguish the ultimate luminosity source.  If these ULIRGs are
powered by compact starbursts, the star formation rates seen in the
central few hundred parsecs far exceed the global rates seen in
nearby starburst galaxies, and approach the surface brightness of
individual clusters in nearby starburst galaxies.

\end{abstract}
 
\newpage

\section{Introduction}

Ultraluminous Infrared Galaxies (ULIRGs; defined as
those systems with  luminosity from 8 - 1000 $\mu$m L$_{8-1000\mu m}\ge
10^{12}L_{\odot}$) were discovered to be a significant class of
objects in the local Universe in the IRAS all sky survey (Soifer et
al. 1987, Sanders et al. 1988).  These systems have recently been
suggested as a major component in the Universe at z$>$ 2 through deep
submillimeter surveys  ( c.f. Lilly et al. 1999,  Blain et al. 1999,
 Barger et al. 1998).  

Whether ULIRGs are powered by dust enshrouded AGN  or starbursts has
been debated since their discovery(e.g. Sanders 1999, Joseph 1999).
This question is made all the more important by the realization that
ULIRGs could produce a major fraction of the radiant energy in
the Universe (Lilly et al. 1999, Heckmann 1998).

A wide variety of observational techniques have been employed to probe
the nature of the central luminosity source in ULIRGS, ranging from
optical spectroscopy (Sanders et al. 1988) to cm radio continuum
imaging (Condon et al. 1991, hereafter C91), to VLBI observations
(Smith, Lonsdale and Lonsdale 1998) and hard X-ray observations (Rieke
1988, Nakagawa et al. 1998, Isawawa 1999).  The nuclear environments
of  these galaxies have been effectively imaged with high spatial
resolution in the optical and near infrared (Surace et al. 1998,
Surace and Sanders 1999, Scoville et al. 1999, hereafter S99).

Observations in the
thermal infrared  have the potential to address the question of the
origin of  the luminosity by establishing the size of the regions that
are radiating  the infrared luminosity.  The  IRAS observations showed
that the mid and far infrared luminosity in these systems emerging 
predominantly  between 50 and 200
$\mu$m was due to
thermal reradiation by dust. However, the spatial resolution of those 
observations,  $\sim
1'-2'$, was inadequate to place meaningful size constraints on the
luminosity sources.  Because the mid-infrared wavelengths (8--25$\mu$m)
carry a
significant fraction of the total bolometric luminosity in ULIRGs,
ranging from $\sim$10\% to $>$ 30\% of the 8--1000$\mu$m luminosity,
 observations at these wavelengths with a resolution $\le 1''$  have the
potential to constrain the sizes of the emitting regions and thereby
address the nature of the underlying sources that heat the radiating
dust.  Observations  in the 10$\mu$m atmospheric window, with
a resolution $\sim 1''$, have been reported for  several ULIRGs by
Miles et al. (1996), and Keto et al. (1992).  In most cases they found the
sources to be unresolved implying that the sources were either very
compact starbursts or dust enshrouded AGN.

With a diffraction limit  of  0.24$''$ FWHM at 10$\mu$m, the Keck
Telescope provides a substantial improvement in spatial resolution over
previous mid-infrared observations,  probing the
distribution of the thermal emission  at the 100-300 pc
scale in nearby ULIRGs.  With the introduction of mid-infrared imaging 
on the Keck
Telescope, we have started a program to address this problem.  In
the first paper from this program we reported imaging
of the nucleus of the closest ULIRG - Arp 220 (Soifer et al. 1999).
The observations were able to spatially resolve the emission at
24.5$\mu$m, and demonstrated that the central thermal source is
optically thick at this wavelength. These observations placed
severe constraints on the surface brightness in the
infrared luminous nuclei.

In this paper we report imaging observations from 8 - 24.5$\mu$m of a
sample of the closest  ULIRGs at spatial resolutions of
0.3-0.6$''$.  These data provide  the highest spatial
resolution yet achieved on the thermal emission from the closest
ULIRGs and trace the spatial distribution of the emergent luminosity
in these systems.   We adopt H$_o$ =75 km s$^{-1}$Mpc$^{-1}$.

\section {The Sample}

The objects observed were taken from the IRAS Bright Galaxy Sample
(Soifer et al. 1987) with the addition of  IRAS 17208-0014 (Sanders
et al. 1995), which is
at a low Galactic latitude.  The basic information for the seven
objects in the sample is given in Table 1.   The objects selected for
observation were chosen on the basis of 1) meeting the luminosity
definition of the ULIRGs, i.e. $L_{8-1000\mu m}$ $\ge
10^{12}L_{\odot}$, 2) having a  25$\mu$m flux density $>$ 1 Jy based
on IRAS measurements,  3) being at the lowest possible redshift for
the best linear resolution, and 4) being available at the time of the
observations.  This sample contains the five closest  ULIRGs in the
IRAS Bright Galaxy sample of Soifer et al., and  seven of the
ten ULIRGs known with $\delta \ge -30 \deg$ and z\,$\le$\,0.06.  The
projected linear scales of these objects range from 360 pc
arcsec$^{-1}$ to 1200 pc arcsec$^{-1}$, so that the point spread
function/angular resolution of the Keck observations,  0.3$''$ at
12$\mu$m and 0.6$''$ at 24.5$\mu$m provide linear resolutions of
100~pc - 700~pc, depending on the object and the wavelength observed.
Three of the objects in the sample have been classified as Seyfert
nuclei, three as LINERS, and one as an HII region, based on optical
spectroscopy.

\section {Observations and Data Reduction}

The observations were made primarily using the MIRLIN mid-infrared
camera (Ressler et al. 1994) at the f/40 bent Cassegrain visitor port
of the Keck II Telescope. The camera uses a 128$\times$128 Si:As array
with a plate scale of 0.138$''$/pixel for a total field of view of
17$'' \times$17$''$.  Table 2 reports the central wavelengths for the
filters through which each object was observed. At each wavelength the
observing procedure was the same.  A secondary with a square wave chop
of  amplitude 6$''$ in the north-south  or east-west direction at 4 Hz
was employed for fast beam switching.  The frames sampling  each chop
position were coadded separately in hardware, resulting in two images.
After an interval of approximately a minute, the telescope was nodded
perpendicular to the chop direction  (east-west or north-south) by
6$''$ and a second pair of images was obtained in order to cancel
residuals in the sky and to subtract telescope emission.  This
procedure was repeated a number of times at each  wavelength.  The
data were reduced by differencing the two images obtained within the
chop pairs at each nod location, and then coadding the resulting
positive images, with the positions appropriately adjusted to a common
location, to yield a positive image centered in a field approximately
6$'' \times$6$''$.  Because of the  chopper and telescope nod spacings
employed for the observations, the  data are not capable of measuring
flux outside a 6$''$ diameter region.

Observations with MIRLIN of all the ULIRGs except IRAS~05189-2524
were obtained in two nights in March 1998.  The observations of the
targets were interleaved with  observations of nearby bright stars
that were used  to establish the point spread function (PSF) for the
observations.  The PSF stars were observed only three times per night
in  each of the bands at 12.5, 17.9 and 24.5~$\mu$m, but an attempt
was made to accompany the critical observations with measurements of a
nearby PSF star.  At 24.5~$\mu$m the PSF was always equal to  the
diffraction limit of the telescope to be 0.62$''$ FWHM.  At the
shorter wavelengths, the size of the PSF image was set by diffraction
and the pixel sampling about half of the time; at other times
atmospheric seeing affected the image size.

IRAS~05189-2524 was observed in  October 1998. During this run the PSF
stars were observed more frequently, but for  wavelengths shorter
than $\lambda \sim$ 17.9~$\mu$m significant variation in the measured
size of the PSF star was seen.

The variation in the measured size of the PSF calibration star is  the
largest uncertainty in the measurement of the source sizes; this issue
is addressed when observations of a particular object are described.

The MIRLIN observations were made under photometric conditions.   The
photometry  was calibrated based on observations of four bright stars,
$\alpha$ Tau, $\alpha$ Boo, $\beta$ Peg, and $\alpha$ Cet whose
magnitudes, in turn, were based on IRAS photometry.  The uncertainties
in the photometry, based on the internal consistency of the
observations, is estimated to be 5\% at $\lambda \le$17.9~$\mu$m and
10\% at 24.5~$\mu$m. The flux density corresponding to 0.0 mag
(Vega-based)  was taken to follow the prescription given in the
Explanatory Supplement to the IRAS Catalogs and Atlases (Joint IRAS
Science Team, 1989), and is given in Soifer et al. (1999).

In addition to the MIRLIN observations, UGC~5101,  Markarian 231 and
IRAS~08572+3915 were observed using the Long Wavelength Spectrograph
on the Keck~I Telescope on the night of 01 April 1999. The night was
not photometric and so only measurements of the sizes of the compact
sources were made; all the observations were carried out at
12.5~$\mu$m. The chopper was set to an amplitude of 5$''$ chopping
north-south at a  frequency $\sim$ 5 Hz. Observations were made  in a
fashion similar to  the MIRLIN observations, except that the telescope
nodding was in  the same direction as the chopping and the nodding
amplitude was the same as the chopping amplitude. The data were
reduced in a manner similar to the MIRLIN data.  The pixel scale is
0.08$''$ pixel$^{-1}$.

Observations of Markarian 273 were made at 3.45~$\mu$m using the
near-infrared camera at the Cassegrain focus of the 200-inch Hale
Telescope at Palomar Observatory in  May 1999.   The camera was read
out in a  64 $\times$ 64 pixel subarray because of the high read rate
required to avoid saturation.  The scale is 0.125$''$
pixel$^{-1}$. The chopper throw was 15$''$ at a frequency of $\sim$1
Hz. Observations of stars given by Elias et al. (1982) were used to
establish the photometric scale.

\section {Results}

The mid-infrared  observations of the ULIRGs reported here  show a
variety of characteristics,  generally consistent with their
properties seen at high spatial resolution at radio wavelengths
(c.f. C91)  and in the near infrared using NICMOS on HST (Scoville et
al. 1998, S99).    The mid-infrared data on all the ULIRGs are
presented in Figures 1, 2 and 3. In Figure 1 we present the
12.5~$\mu$m images of the ULIRGs and, where the radio and near
infrared data are available, compare these maps to the equivalent maps
at 8.4~GHz  from the data of C91 and to the NICMOS 2.2~$\mu$m data
from S99 and Scoville et al (1998). In the case of three galaxies,
IRAS~05189-2524, IRAS~08572+3915 and Markarian 231, where there is a
single compact nucleus that emits a substantial fraction of the
12.5$\mu$m flux density, we have presented the corresponding PSF star
at 12.5$\mu$m for comparison. In Figure 2 we compare,   for the five
single nucleus sources, the integrated flux at 12.5 $\mu$m as a
function of beam radius with the same quantity for a  corresponding
point source to establish the spatial distribution of the mid-infrared
radiation.  Figure 3 displays the spectral energy distributions (SEDs)
obtained from the mid-infrared data and compares these data with the
corresponding IRAS data and data at other wavelengths.  The flux
densities measured for the ULIRGs in a 4$''$ diameter beam are
presented in Table 3.
 
The mid-infrared observations did not establish  accurate absolute
positions for the sources. In some cases, described below, we use
arguments regarding the similarities of the structures seen at vastly
different wavelengths to argue for the physical coincidence of the
sources seen at different wavelengths.

In the following sections we give results for the individual objects.

\subsection {IRAS~05189$-$2524}

MIRLIN observations of IRAS~05189$-$2524 were made at 12.5 and 24.5
$\mu$m.  The shorter wavelength data have the higher signal-to-noise
ratio, and so are  presented in Figure 1.  Images at 12.5$\mu$m (MIRLIN),
2.2$\mu$m (NICMOS, S99) and  8.4GHz (VLA, C91) are shown in Figure 1a.
The dominant central source is clear in all three maps.  At 8.4GHz,
the single source is concentrated with a measured size of 0.2$''
\times$0.17$''$(C91).  At 2.2~$\mu$m, the unresolved source is
$<0.2''$ and  contains $\sim$75\% of the total flux measured in a
5$''$ diameter beam, and 70\% of the flux contained in an 11$''$
diameter beam (S99).

The source appears virtually unresolved in the MIRLIN image.  As can
be seen from Figure 2, there is excellent agreement between the curve
of growth for IRAS~05189-2524 and the PSF star observation.   The
apparent size of the PSF star varied  during the time of the
observations, so the limit on the size of IRAS~05189-2524 is not as
stringent as would otherwise be achieved.   At 12.5$\mu$m the  size is
$<$0.2$''$ (Table 4). The source size at 24.5~$\mu$m is an upper limit
as well, and is larger and thus consistent with that established at 
12.5~$\mu$m.

As seen in Figure 3 and Table 3, the flux densities measured by MIRLIN at
12.5~$\mu$m and 24.5~$\mu$m agree well with the IRAS observations of
this source; virtually all of the luminosity of this source at
12.5~$\mu$m and 24.5~$\mu$m is contained within the unresolved source
measured here.

The MIRLIN observations of IRAS~05189$-$2524  lead to a limit on
the source diameter of $<$170 pc in the mid-infrared (Table 5).  The
combination of flux density and upper limit on the source sizes(Table 4), 
imply the 
brightness temperatures T$_b>$123~K and $>$85~K at  12.5~$\mu$m and
24.5~$\mu$m respectively.  The flux ratio between   12.5~$\mu$m and
24.5~$\mu$m corresponds to a color temperature T$_c =$169~K, while the
color temperature derived from the 25$\mu$m and 60$\mu$m IRAS
measurements is 83~K\footnotemark[1]. 

\footnotetext[1]{The brightness temperature, T$_b$, is defined by
$f_{\nu} = \Omega \times B_{\nu}(T_b)$
where $B_{\nu}(T_b)$ is the blackbody emission at frequency $\nu$ for
temperature T$_b$ and $\Omega$ is the solid angle subtended by the source. 
The color temperature of a source, T$_c$, measured between frequencies $\nu_1$
and $\nu_2$ is found from 
$\frac{f_{\nu_1}}{f_{\nu_2}}$ $=$ $\frac{ B_{\nu_1}(T_c)}{B_{\nu_2}(T_c)}$.}

A central  source with  the bolometric luminosity of IRAS 05189$-$2524
would heat silicate or graphite grains to the apparent color
temperature between 12 and 24.5$\mu$m, T$_c\sim$ 170~K,  at a radius of
30-60 pc using the grain parameters of Draine and Lee (1984). This is
much smaller than the observational limit. Alternatively, the   size
of an optically thin source  which  produces the 25$-$60 $\mu$m flux 
ratio with a color
temperature of 83 K is significantly larger, 170-250 pc in radius,
again using the Draine and Lee grain parameters.

\subsection {IRAS~08572+3915}

MIRLIN observations of IRAS 08572+3915 were obtained at five
wavelengths (Table 2). In addition, LWS observations of
IRAS~08572+3915 were obtained at 12.5~$\mu$m.  Optical and near
infrared imaging (Sanders et al. 1988, Carico et al. 1990, S99) show
two galaxies separated by about 5$''$ in a northwest-southeast orientation.  
The southeast
source is not detected in the radio (C91) or in  the present data.

The 12.5$\mu$m LWS image in Figure 1b shows a bright central source
(the northwest source in the near infrared) that is unresolved.  The
size limit at 17.9~$\mu$m is consistent with that found at
12.5~$\mu$m.  Figure 1b and Figure 2 show  that the central source is
dominant in the mid-infrared and radio, containing $\sim$80\% of the
total flux measured in a 4$''$ diameter beam.  At 2.2~$\mu$m, the
central unresolved source contains $\sim$60\% of the total  flux in an
11.4$''$ diameter beam (S99).

The most stringent  observations constraining the size and spatial
distribution of the  source are the LWS observations at 12.5$\mu$m.
Five observations  of the PSF star (BS3275) were interleaved with
eight observations of the object. The average FWHM of the object (0.39
$\pm$ 0.04$''$) was marginally greater than the average FWHM of the
PSF  (0.34 $\pm$ 0.01$''$).   Formally this leads to  a size of the
12.5$\mu$m source of 0.19$''$, but this is only a $1 \sigma$ result.
A $3 \sigma$ limit on the  size of the compact source is
$\sim$0.3$''$. As illustrated in Figure 2, there appears to be a
measureable flux, approximately 20\% of the total at 12.5~$\mu$m, that
extends beyond the central core, and  is contained within 4$''$.  The
figure shows clearly that the flux measured in IRAS~08572+3915
continues to increase beyond the beam where the flux in the PSF star
has stopped increasing.

Figure 3 and Table 3 compare the Keck photometry in a 4$''$  diameter
beam for IRAS~08572+3915 with the IRAS measurements.  The MIRLIN
photometry accounts  for 70\% of the IRAS flux density at 12.5$\mu$m.
The MIRLIN observations were taken with comparatively narrow filters,
i.e., $\Delta\lambda$/$\lambda < \sim$ 0.1,  from 7.9$\mu$m to
17.9$\mu$m, while the IRAS 12$\mu$m filter bandpass extends from
8-15$\mu$m. Given the uncertainties in the corrections for the
different bandpasses, the MIRLIN data are consistent with all the flux
detected by IRAS being confined to a 4$''$ diameter beam surrounding
the  detected source although it is possible that some of the flux
could be accounted for by extended emission associated with the
northwest source or from the southeastern source.

Neither the  MIRLIN nor the  LWS observations detected the galaxy
approximately 5$''$ southeast of the object centered in  Figure 1b.
The MIRLIN data limit the contribution of the southeastern source to
$<$30\% of the total 12.5$\mu$m flux; it is likely that  the
southeast galaxy  does not contribute significantly to the total
infrared luminosity in this system. This is consistent with the
optical and near infrared morphology and colors of this galaxy, which
are  indicative of quiescent galaxy nucleus (Surace et al. 1998, Surace
and Sanders 1999).

\subsection {UGC~5101 = IRAS~09321+6134}

MIRLIN observations of UGC~5101 were obtained at 12.5 and 17.9~$\mu$m.
Contour plots of this source in Figure 1c show that  at all wavelengths
the structure is basically the same: a bright central core centered on
an apparently elliptical structure having a major axis at a position angle of
$\sim$80$^{\circ}$.  While the position angle of the elliptical structures
agree well at 2.2$\mu$m and 8.4 Ghz, there is a suggestion that the 
position angle of
the ellipse at 12.5~$\mu$m is $\sim$15$^{\circ}$ smaller.
At 8.4 Ghz the central source contains 80\% of the radio flux(C91),
while at 2.2$\mu$m, the unresolved peak contains 35\% of the flux in a
5$''$ diameter beam (S99).

As shown in Figure 2, at 12.5~$\mu$m approximately half of the total
flux measured in a 4$''$ diameter beam is contained in the central
compact source. The inner central compact source is indistinguishable
from the PSF (Figure 2). From the dispersion in the sizes determined
for the PSF star (BS3775) the compact source has a size (3$\sigma$)
less that 0.22$''$. The size measured at 17.9$\mu$m is consistent with
that measured at 12.5~$\mu$m.

The data of Figure 3 and Table 3 show that at 12.5~$\mu$m the
flux measured by MIRLIN is $\sim$60\% of that measured by IRAS.  This
suggests that there is low surface brightness emission extended on a
larger scale than 4$''$. The 12.5$\mu$m flux density contained in the
compact source is $\sim$30\% of the  total flux density at 12$\mu$m 
measured by IRAS.

\subsection {Markarian 231 = UGC~8058 = IRAS~12541+5708}

Markarian 231 has long been known both as an infrared luminous system
(Rieke and Low, 1972) and a Seyfert 1 system (Boksenberg et al. 1977). The IRAS
survey found that it is the most luminous object within 300 Mpc, with
a bolometric luminosity of $3\times 10^{12}L_{\odot}$.

MIRLIN images of Markarian 231 were obtained at seven wavelengths from
7.9 to 17.9$\mu$m. In addition, LWS imaging was obtained at
12.5$\mu$m.  Figure 1d shows contour plots of Markarian 231 at 12.5
$\mu$m from the LWS data and data at 8.4 Ghz(C91).   The emission at
all the mid-infrared wavelengths is dominated  by the very bright
central source, and is effectively unresolved.  The radio continuum is
 entirely from the bright nucleus.

LWS imaging was used to establish the size of the central source at
12.5$\mu$m. The LWS observations consisted  of eleven measurements of
the size of Markarian 231, interleaved with nine measurements of a
near-by PSF star(BS 4905) over a period of two hours.  The average
FWHM  of the PSF star  and Markarian 231 were indistinguishable and
were  slightly greater than the diffraction limit of the telescope and
the rms (population) scatter in the FWHM was 0.04$''$.   The central
source thus is  unresolved in these observations. We have chosen to
set a 3$\sigma$ limit on its FWHM of 0.13$''$ on the basis that a
difference three times the average standard deviation of the FWHM
would have been easily detected.

The SED of Markarian 231 is plotted in Figure 3, and shows clearly
that all the flux measured by IRAS is detected in the 4$''$ diameter
beam.  The SED shows a dip at 10$\mu$m  apparently due to foreground
silicate absorption of the central source.

\subsection {Markarian 273 = UGC~8696 = IRAS 13428+5608}

The MIRLIN  observations of  Markarian 273 consist of imaging at seven
wavelengths from 7.9$\mu$m to 17.9$\mu$m. Figure 1e shows a
comparision of the MIRLIN 12.5$\mu$m data with the 8.4~GHz map (C91)
and  the NICMOS 2.2$\mu$m image (S99), while a montage of images of
Markarian 273 is shown in Figure 4.  At 12.5$\mu$m the source
structure is basically double, with the northern source brighter than
the southwestern source at most wavelengths.  A similar morphology is
seen at 2.2$\mu$m (S99), but not in the radio (C91).  In comparing the
12.5$\mu$m, NICMOS and 8.4 Ghz data we have assumed the spatial
coincidence between the bright northern source in each image.  The
overall ageement between the 12.5$\mu$m and NICMOS images is
excellent, while the discrepancy between the positions of the
southwestern source in the infared and radio is real.  The position
angle between the northern and southwestern sources is the same (to
within the uncertainties) in the 12.5$\mu$m, 2.2$\mu$m and 8.4 GHz
images, while the source separations  are 1.0$''$, 1.0$''$ and 1.2$''$
at these same wavelengths.  The apparent separation between the
infrared and radio sources, 0.2$''$, corresponds to 160pc at the
distance of Markarian 273.  Note also that the flux ratio between the
northern and southwestern source is considerably different between the
infrared and radio.

As shown in Figure 4, the structure seen in Markarian 273 at all
infrared wavelengths from 3.4$\mu$m to 17.9$\mu$m is  basically the
same.  There is no evidence  for a mid-infrared counterpart to the
southeastern source seen at 8.4 Ghz by  C91 (Figure 1e) and at 5 Ghz
by Knapen et al. (1997). The bright source present to the southeast in
the radio map is suggested to be a background radio object by Knapen
et al., although the statistical likelihood of a background
source falling within 1$''$ of the location of the radio core of
Markarian 273 is $< 10^{-5}$.  There is a faint compact object found
in the NICMOS images (S99) at the location of southeast radio source,
although there is no evidence for any mid-infrared emission at this
location. The upper limit on the 12.5$\mu$m flux density at the
location of the southeastern radio source of 15 mJy is  1/10 that of
the  northern source.  This limit, coupled with the 8.4GHz flux
density of this source of 3.2 mJy (C91), leads to a limit on the ratio
of flux densities $\frac{f_{\nu}(12.5\mu m)}{f_{\nu}(8.4 GHz)}$ $<$
4.7, consistent with the observed ratio $\frac{f_{\nu}(12.5\mu
m)}{f_{\nu}(8.4 GHz)}$ $=$4.5 for  the northern source.  By
comparison, if the radio ( 0.2mJy at 8.4 GHz, C91, Knapen et
al.) and infrared (58 mJy at 12.5$\mu$m) flux densities of the
southwestern source are assumed to be associated with the same
physical source, the ratio $\frac{f_{\nu}(12.5\mu m)}{f_{\nu}(8.4
GHz)}$ $=$ 290, a factor of 65 times greater than that seen in the
northern source, making this nucleus far more infrared bright relative
to the radio than the northern source.  If the southwestern radio and
infrared sources are not related, the southwestern infrared  source
has an even larger $\frac{f_{\nu}(12.5\mu m)}{f_{\nu}(8.4 GHz)}$
ratio.

As can be seen in Figures 1e and 4, the relative brightnesses of the
nuclei  change substantially as a function of wavelength. In order to
get an estimate of the relative flux emitted by the two nuclei, the
flux in two identically sized  square beams was measured at the same
locations on the images at all wavelengths presented; the beams,
1.6$''$ on  a side,  are shown in Figure~4.  The sky was taken
northwest and southeast of each of the squares. The efficiency of the
square beams was estimated by placing the PSF of the appropriate
wavelength at the locations of the nuclei within the square beam. The
photometry of the NICMOS image was arbitrarily normalized so that the
sum of the fluxes from the two squares equals the sum of the fluxes
give by Knapen et al. (1997) and the flux density in a 4$''$ diameter
beam equals that of Knapen et al. The results of these photometric
measurments are shown in Figure 5.

At 3.4$\mu$m and 10.3$\mu$m  the southwestern source is  slightly
(10\%) brighter than the northern  source.  At the other mid-infrared
wavelengths the northern source is between a factor of 1.5 and 3 times
brighter than the southwestern source.  The different SED's of the
components implied by their changing relative brightnesses is shown
directly in Figure 5, where photometry of the two nuclei is shown.
The northern source clearly shows a signficantly deeper absorption at
10$\mu$m than the southwestern source. This is consistent with the
distribution of CO emission which strongly peaks on the
northern source (Downes and Solomon, 1998), but is
inconsistent with the near infrared colors found by S99, 
where the northern source  has a  smaller J-K color than
does the southwest source.

The SED of Figure 5, particularly the relatively strong emission at
3.5 and 11.7 $\mu$m, suggests aromatic hydrocarbon emission (PAH)
emission associated with the southwest nucleus. Such emission is seen
in the spectra of many ULIRGs (Genzel et al. 1998).  The emission
from the southwest source could  account for all of the 11.3\,$\mu$m
emission seen in the spectrum of the combined nuclei by
Dudley(1999).The northern source does not show a similar  peak at
11.7$\mu$m, but clearly shows the effects of substantial overlaying
cold silicate absorption.

The MIRLIN photometry shown in Figure 3 was obtained using a 4$''$
diameter beam centered between the two nuclei.  The comparison of the
MIRLIN and IRAS data (Figure 3) demonstrates that the mid-infrared
emission in this source emerges virtually entirely from the central
region.  Because of the proximity of the two nuclei and the fairly low
signal-to-noise ratio of the nuclei in some of the images, it is
difficult to cleanly determine the sizes of the two sources. The data
suggest that both nuclei have unresolved mid-infrared cores with sizes
$<$0.4$''$. In the northern nucleus approximately 2/3 of the
12.5$\mu$m flux is unresolved, while in the southwestern nucleus,
about half the flux is unresolved.

\subsection {Arp 220 = UGC~9913 = IRAS 15328+2340}

The data for this source were discussed in detail in Soifer et al. (1999)
and are included in Figures 1f and 3 for completeness.

\subsection {IRAS~17208-0014}

MIRLIN imaging was obtained at 7.9 and 12.5$\mu$m.  The 12.5$\mu$m
image is shown in Figure 1g.   The 2.2$\mu$m NICMOS image of
IRAS~17208-0014 (Figure 1g, S99) shows it to be an extended source of
roughly 2$''$ diameter with a number of luminous starburst centers to
the northeast of the nucleus.  The mid-infrared image is apparently
also similar, but the limitation due to the chop-nod spacing and the
relatively low infrared surface brightness results in an image with
low signal-to-noise ratio. The size of the 12.5\,$\mu$m source of
$\sim$2$''$ or 1.6 kpc FWHM (Figure 2) is quite similar  as well to that
found in CO by Downs and Solomon (1998) and Jogee et al.(2000).

The best illustration of the spatial distribution of the mid-infrared
emission is shown  in Figure 2. The source flux
smoothly increases  up to 5$''$ diameter, with a FWHM of 2$''$; the
apparent leveling off of the enclosed flux at a diamteter of 5$''$ is
caused by the artificial imposition of the sky level between diameters
of 5 and 6$''$.
 
Evidence that the source is not extended much beyond a diameter of
5$''$ is provided by Figure 3 which shows that,  within the
uncertainties, the enclosed flux within a 4$''$ diameter beam is the
same as within the IRAS beam.  The lack of a dominant compact
component in the mid-infrared make this object very different from the
other ULIRGs in this sample.

The radio emission from IRAS~17208-0014 is much more compact than the
infrared emission.  At least 90\% of the 1.4GHz flux comes from a
region smaller than 2$'' \times$2$''$ (FWHM, Condon et al. 1996).
Martin et al. (1989) measure a radio source with a  size of 0.26$''
\times$0.32$''$ at 1.6 GHz that accounts for $\sim$90\% of the 1.4 GHz
flux density, suggesting a compact radio core.  This core is clearly
not dominant in the mid-infrared imaging.  The near infrared NICMOS
imaging of this galaxy (S99) suggests that the nucleus could be
obscured at 2.2$\mu$m, but the lack of detection of a strong compact
core at 12.5$\mu$m is quite peculiar, since the radio source is so
compact.

\section {Discussion}

\subsection {Compact  Sources in ULIRGs}

The data presented  here, with angular resolution of $<$0.3$''$,
represent the highest spatial resolution achieved to date  in probing
the central regions of ULIRGs in the thermal infrared.   Table 4
summarizes the key observations, the sizes of the compact sources
found in the ULIRGs at 12.5\,$\mu$m,  17.9\,$\mu$m and 24.5\,$\mu$m,
as  well as the fraction of the total flux at these wavelengths
emitted from these cores.  All of the ULIRGs except IRAS~ 17208$-$0014
have a  nuclear source in the mid-infrared of size $\leq$200 pc that
produces a substantial fraction, if not all, of the galaxy's
mid-infrared luminosity.  In three sources, IRAS 05189$-$2524, IRAS
08572+3915 and Markarian 231, a single unresolved nucleus produces
between 80\% and 100\% of the total mid-infrared luminosity.    In
Markarian 273 and Arp 220, two nuclei produce nearly all the
mid-infrared emission in the galaxies. In the case  of Markarian 273
the nuclei are substantially unresolved, while in Arp 220 at least 
one  nucleus is
resolved. Because of its low redshift the linear size of the
resolved nucleus in Arp 220 is smaller than the limit  on all the unresolved
nuclei  except Markarian 231(see Table 5).  In only  one compact
system,  UGC 5101, does the  compact nucleus produce significantly
less than half the total mid-infrared emission from the system. 

As
noted previously, IRAS~ 17208$-$0014 presents an interesting
puzzle. There is no mid-infrared evidence for a compact nuclear
source, but the radio data (C91, Martin et al. 1989) strongly
suggest a power source with a size $<$0.3$''$.  In the following
discussion, we will adopt the morphology implied by the 12.5$\mu$m
image, but we note that extremely high 12.5$\mu$m extinction could be
leading to an erroneous  interpretation of the nuclear structure in
IRAS~ 17208$-$0014.
 
Table 5 shows that  the compact nuclear sources have linear sizes
typically $<$200 pc in diameter.  In  Arp 220 the diameter of the
western source is 140 pc at 24.5$\mu$m (Soifer et al. 1999), while in
Markarian 231,  the diameter of the 12.5$\mu$m source is $<$100 pc.

\subsection {AGNs as the Luminosity Source in ULIRGs }

One of the goals of the observations reported here is to distinguish the
ultimate  source of the infrared luminosity in these systems by means of
 the
spatial extent  of the infrared emission. If the dust is heated by a bright
compact object such as an AGN, the dust emission will be compact, and
the size of the infrared source will increase with increasing
wavelength roughly as $r \sim \lambda^{2.5}$, since the dust
temperature will decrease with radius as $T\sim r^{-0.4}$ and the wavelength
of maximum emission $\lambda_{max}\sim T^{-1}$.  If, on the other
hand, the heating is due to many hot stars in a starburst region, then
the dust emission will be distributed on the same scale as the
young, luminous stars.   Since the dust heating is due to the  proximity of the
dust to the luminous young stars there will be little or no temperature
gradient seen.  Spatially extended 10~$\mu$m
emission therefore implies starburst activity, while spatially compact
10~$\mu$m emission could be due to either a compact starburst or an
AGN.

To quantify the compact mid-infrared emission we utilize the
brightness temperatures derived from our observations and the color
temperatures determined from the IRAS 25 and 60$\mu$m observations 
of these objects.
We assume that T$_c$  measures the temperature of the grains heated by
the luminosity source; T$_b$ should  be $\le$T$_c$ if the
optical depth $\tau_{\nu}<1$.  Only in the limit of large optical
depth should the brightness temperature approach the color temperature.  

Based on the MIRLIN observations of Arp 220, Soifer et al. (1999)
showed that the dominant luminosity source, Arp 220W,  is optically
thick with a brightness temperature at 25~$\mu$m equal to the color
temperature between 25 and 60~$\mu$m.  The observations reported here
permit analogous constraints to be placed on some of the nuclear
sources in other ULIRGs in this sample.  In Markarian 231, IRAS
05189$-$2524 and  IRAS 08572+3915, the limits on the sizes of the
central infrared sources are  sufficiently small that it is worthwhile
considering how these sizes  constrain the far infrared sources.

The size limit on  Markarian 231 of 0.13$''$ at 12.5$\mu$m places the
most severe limits on models of the  infrared emitting region.  In
this case the observed flux density and size limit lead to a lower
limit on the brightness temperature of T$_b \ge$~141K at 12.5$\mu$m.
If the infrared emitting source has the same size at 17.9~$\mu$m,
25~$\mu$m and 60~$\mu$m, the corresponding brightness temperatures
are 124~K, 128~K and 241~K. The brightness temperatures at 12.5$\mu$m,
17.9$\mu$m and 25~$\mu$m are consistent with each other and with a
nearly optically thick source over this wavelength range, but the
60~$\mu$m brightness temperature is much too large to be  consistent
with dust  models with reasonable grain properties. This implies that
the 60$\mu$m source size is much  larger than the 12.5$\mu$m size.

The color temperature measured between  25 and 60 $\mu$m, determined
from the IRAS measurements, is 85~K. The size of a blackbody of 85~K
required to produce the observed 60$\mu$m flux density is 0.4$''$,
significantly larger than the  limit on the source size at 12.5$\mu$m.
The source size determined if T$_c =$~T$_b$ represents a lower limit
on the size of the 60$\mu$m source.    For example, if the size of the
60$\mu$m source were equal to the $\sim$1$''$ size found in CO by
Downs and Solomon (1998), the 60$\mu$m brightness temperature would be
53K, much less than that inferred by assuming all the 25 and 60 $\mu$m
emission is emerging from the same region.    The size limits are
based  on the assumption of  large optical depth ($\tau >>$1); smaller
optical depth would imply an even  larger  source.   Since significant
emission is seen at $\lambda < 60 \mu$m, either $\tau_{60 \mu m}<$1 or
we are viewing a tilted optically thick disk, as suggested for Arp 220
(Scoville et al. 1998, Downs and Solomon, 1998,  Sakamoto et al. 1999,
Soifer et al. 1999).  In either case, the size constraints  show that
in Markarian 231 the infrared source increases in size with increasing
wavelength.  This is consistent with the picture of an AGN   with
luminosity of 3$\times$10$^{12}$ L$_{\odot}$  heating the surrounding
dust and producing a radial temperature gradient.

In this sample there are two additional ULIRGs where significantly
more than 50\% of the mid-infrared luminosity emerges from a compact
central source, IRAS~05189-2524 and IRAS~08572+3915.  In
IRAS~05189-2524, if the size of the 60$\mu$m emitting source is the
same as the limit on the 12.5$\mu$m source measured here,
T$_b$(60$\mu$m)=101~K, which is significantly greater than  T$_c=$85~K
from the IRAS 25 and 60$\mu$m data. For the same assumptions, in
IRAS~08572+3915  T$_b$(60$\mu$m)=77~K is slightly less than T$_c=$83~K
determined from the IRAS data at  25 and 60$\mu$m. For both sources,
using dust properties described by Draine and Lee (1984) the sizes
calculated for silicate and graphite dust grains heated by central
sources of the observed bolometric luminosity, and having dust
temperatures equal to the  observed color temperatures are larger by a
factor of 1.5 to 3 than the observed sizes of these sources at
12.5$\mu$m.  This suggests that the sizes of these infrared sources
also increase with increasing mid and far-infrared wavelength.

High apparent surface brightnesses and  a radial temperature gradient
are  natural consequences of centrally heated dust emission
surrounding a high luminosity source, i.e., an AGN. In Markarian 231
there is direct evidence that the size of the infrared source
increases with increasing wavelength in the mid and far infrared. In
IRAS~05189-2524 and  IRAS~08572+3915, there is strong evidence for
such radial temperature gradients based on the compact sources
observed in the mid-infrared.  The observed luminosities are adequate to
quantitatively account for the inferred variation of dust temperature
with radius if the dust is heated by a central source in these objects.

\subsection {Starbursts as the Luminosity Source in ULIRGs and 
Energy Production Rates}

The compact nature of the sources in the objects observed leads to
very high luminosities produced in small volumes.  By comparing these
energy production rates with those found in  local  examples of
starburst environments, we can  determine whether known starburst
environments can generate the apparent  energy production in the
ULIRGs.

In the core of Orion the luminosity of $2\times 10^5 L_{\odot}$
emerges from a  0.3 pc size region (Werner et al. 1976), while in the
nuclear starburst in M82, the nearest starburst galaxy, the luminosity
of $3\times 10^{10} L_{\odot}$ is emitted over a projected area of
450$\times$75pc (30$'' \times$5$''$) (Rieke et al. 1980).  These
observations lead to apparent surface brightnesses of emergent
luminosity of $1-3\times 10^{12}~L_{\odot}$/kpc$^2$. If, as seems
likely, M82 is being viewed edge-on, then the mean surface brightness
of this  system, if viewed face on,  is closer to 2$\times
10^{11}~L_{\odot}$/kpc$^2$.
  
A survey of the maximum surface brightness in nearby starburst
galaxies in the UV, infrared and radio by Meurer et al. (1997) leads to
peak global averaged surface brightnesses in  starburst galaxies  of
$\sim2\times 10^{11}~L_{\odot}$/kpc$^2$.  Meurer et al.  find that the
peak surface brightnesses of clusters in starburst galaxies reaches $\sim
5\times 10^{13} L_{\odot}$/kpc$^2$ over regions $\sim$10 pc  in
diameter.  Meurer et al. suggest that the global peak surface
brightnesses  is due to a mechanism whereby the pressure from stellar
winds and supernovae is able to regulate the star formation rates in
galaxy disks, while the formation  of massive stars in giant HII
regions is regulated by the ``Kennicutt Law'', i.e. star formation
occurs when the surface gas density exceeds a critical value that
depends on the rotational velocity and shear (Kennicutt, 1989).

The surface brightnesses observed in the ULIRGs can be estimated
if we assume that the size (or limit) of the region measured in the
mid-infrared is representative of the infrared emitting region as a
whole.  With this assumption, we have presented  the surface
brightnesses for the observed ULIRGs in Table 5. We include two values
of the surface brightness, a ``maximum'' which assumes that the
fraction of the total infrared luminosity emerging from the compact core is
the same as the fraction of the mid-infrared flux density emerging
from the  core, and a ``minimum'' that assumes that only the
{\it observed} mid-infrared  luminosity emerges from the compact core.
In the case of Markarian 231, the former number is 
unphysically large (section 5.2).
In the other systems, the two values represent likely upper and very
conservative lower limits on the apparent surface brightness of the
infrared emission.

Only in  the case of IRAS~17208$-$0014, where the near and
mid-infrared observations  suggest the luminosity is due to star
formation, is the measured infrared surface brightness comparable to
the peak global value  $2\times 10^{11}~L_{\odot}$/kpc$^2$ found by
Meurer et al.(1997).  Other than this case, the inferred surface
brightnesses are much greater than found globally in disk starburst
systems.  The highest apparent surface brightnesses in the ULIRGs are
found  in the western nucleus of Arp 220, which is most likely powered
by a nuclear starburst (Smith et al. 1998b) and the cores of Markarian
231 and IRAS~05189$-$2524, which are probably powered by AGN.  In
these cores the apparent surface brightness approaches or surpasses
the surface brightness of clusters in starburst galaxies, while
in the other systems (excluding IRAS~17208$-$0014) the surface
brightness most probably  exceeds  $10^{13}~L_{\odot}$/kpc$^2$.  In
all these cases the surface brightness over regions of hundreds
of parsecs across are comparable to the brightness found in
$\sim$10 pc sized clusters within starburst systems by Meurer et al.

For the ULIRGs that are powered by star formation, the large surface
brightnesses and sizes suggest that the global starburst intensity
regulating mechanism of Meurer et al. (1997) does not apply.
Gravitational instability allows for higher gas densities and hence
higher star-formation rates in the solid-body (no shear) rotation
nuclear regions.  Since the nuclei of most ULIRGs are mergers, far
from equilibrium, the gas  densities may be very high, leading to high
star formation rates and correspondingly high surface brightnesses for
short times.  ULIRGs are clearly  extraordinary events in the
evolution of galaxies. Several  studies (e.g., Sanders et al. 1988,
Scoville, Yun and Bryant 1997, Downes and Solomon 1998) suggest
that the nuclear disks are unstable in some ULIRGs.  The large gas
masses in these environments might then lead to nuclear star formation
rates  much greater than those found in less extreme systems.

\subsection {A Panchromatic View of ULIRGs}

Table 6 summarizes the characterization of the ULIRGs applying many
different diagnositics of the underlying energy sources. Here we have
summarized the structure in the mid-ir from the data presented here,
in the near infrared from S99 and Surace and Sanders (1999), and in the radio
from the VLBI data of Smith et al. (1998). The
optical spectral classifications are  from Sanders et al. (1988) and
Veilleux et al (1995).   The  near-infrared spectral classifications are
from Veilleux et al. (1999) and Murphy (private communication). The ISO
spectral classifications are based on  the strength of the PAH
features and are taken from Genzel et al. (1998) and  Lutz, Veilleux and
Genzel (1999). The X-ray classifications are from ASCA results and are
taken from Iwasawa (1999) and Nakagawa et al. (1999).

In none of the objects in this sample, excepting perhaps Markarian 273
and IRAS17208-0014, is there a completely consistent set of
``classifications'' based on the wide range of wavelengths over which
these objects are observed. In the object that is the clearest case
for a  central AGN luminosity source, Markarian 231, the hard X-ray
emission is not consistent with the simplest view of such power
sources (Nakagawa et al. 1999).   In the  predominantly Seyfert
nucleus IRAS~05189-2524 there is not a dominant compact radio core.
In Markarian 273 the two nuclei clearly seen in high resolution
imaging  complicates the classification, since none of the spectral
classifications resolved the separate nuclei. In UGC 5101, another
source where the optical, infrared and X-ray  indications point towards
a starburst power source, a compact radio core characteristic of a
central AGN exists.  Compact mid-infrared sources occur in six of
seven of these sources, where other evidence points strongly to both
AGN and starbursts as the dominant underlying power source.  Table 6
shows that most ULIRGs do not present a picture over all wavelengths
consistent exclusively with  either an AGN or starburst power source.
In most objects observed here, it is likely that both contribute to
some extent to their total bolometric luminosity, though  Markarian
231 and IRAS~17208-0014 appear to represent the extremes of dominant
AGN and starburst sources.

\section {Summary and Conclusions}

Diffraction limited mid-infrared observations of a sample of seven ULIRGs
obtained on the Keck 10m telescopes have shown:

1 - Extremely compact structures, corresponding to  emission on
spatial scales of $<$100$-$200 pc, are  seen in six of the seven
systems observed. The mid-infrared emission in these galaxies
generally emerges from a single region of diameter $<$ 1 kpc, and in
most cases from regions $<$200 pc in diameter.

2 - The upper limit on the diameter of the 12.5\,$\mu$m source in
Markarian 231, 0.13$''$, is too small to be consistent with the
observed 60\,$\mu$m flux density emerging from a region this size, and
demonstrates that in this AGN the source size increases with increasing
wavelength (decreasing dust temperature).

3 - In IRAS~05189-2524 and IRAS 08572+3915 there is strong circumstantial
evidence that the size of the infrared source increases with wavelength
between 12.5 and 60~$\mu$m, suggesting that these too are predominantly 
AGN powered ULIRGs.

4 - In the other objects observed, the angular size limits for the
compact mid-infrared sources  are not yet adequate to distinguish
between AGN and starburst luminosity sources, although the small sizes
are consistent with a central luminosity source heating the
surrounding dust as in AGN models  in all the sources except
IRAS~17208-0014.

5 - If star formation powers the luminosity in
the very compact infrared sources seen in these systems, the star
formation rates averaged over a few  hundred parsecs is comparable to
that seen in the brightest $\sim$10 pc sized clusters in nearby
starbursts, and  exceeds by factors of up to 100, the  global star
formation rates  in  nearby starburst galaxies.

\section {Acknowledgments}

We thank J.  Aycock and R. Campbell for assistance with the
observations, R. Goodrich,  R. Moskitis and the entire Keck staff for
their help establishing the visitor port and to Barbara Jones, Rick
Puetter and the  Keck team that  brought the LWS into service,
enabling these observations. The W.  M. Keck Observatory is operated
as a scientific partnership between the California Institute of
Technology, the University of California and the National Aeronautics
and Space Administration. It was made possible by the generous
financial support of the W. M. Keck Foundation.  This research has
made use of the NASA/IPAC Extragalactic Database which is operated by
the Jet Propulsion Laboratory, Caltech under contract with NASA.

B.T.S, G.N., K.M. and E.E. are  supported by grants from the NSF and
NASA.  J.A.S. and B.T.S. are supported by the SIRTF Science Center at
Caltech.  SIRTF is carried out at the Jet Propulsion Laboratory.
E.E.B., A.J.W,  N.Z.S.
and  A.S.E. were supported by NASA grant NAG5-3042.
This work was carried out in part (M.W.W., M.R.) at the Jet Propulsion
Laboratory, operated by the California Institute of Technology, under
an agreement with NASA. The development of MIRLIN was suported by
NASA's Office of  Space Science.  The National Radio Astronomy
Observatory (J.J.C.) is a facility of the National Science Foundation
operated under cooperative agreement by Associated Universities, Inc.

\newpage

\figcaption{} Figure 1a. Images of IRAS~05189-2524 obtained at
2.2$\mu$m   (S99, upper left), 8.4 GHz (C91, lower left), and
12.5$\mu$m (this work,  upper right).  The lower right panel is the
PSF star image obtained at 12.5$\mu$m contemporaneously with those 
of IRAS~05189-2524 and used
to determine the  12.5$\mu$m size of  IRAS~05189-2524. The peak of each
image  is at the location 0,0 in each panel.
The contours in
each panel are set so that the peak contour is 90\% of the maximum
surface brightness. The contours are spaced down by a factor of 1.34
so that the third contour down from the peak is at 50\% of the peak
surface brightness.  The ninth coutour down is 8.5\% of the
peak. Below this level the contours are spaced by a factor of two. The
faintest contour is  set at about 3 times the noise in the image. The
mid-infrared data are from the MIRLIN observations. The hatched circle 
show the FWHM of the PSF for the 2.2$\mu$m observations.

Figure 1b. Images of IRAS~08572+3915.  The format, source locations
 and contour levels
are the same as in Figure 1a. The  mid-infrared data are from the LWS
observations. The 2.2$\mu$m and 8.4 GHz data are from S99 and C91
respectively. The hatched circles show the FWHM of the PSF for the
appropriate observations.

Figure 1c. Images of UGC 5101. The format is 2.2$\mu$m image(S99) - left,
12.5$\mu$mimage - center, 8.4 GHz image(C91) - right. The source locations
and contours are set
as in Figure 1a. The  mid-infrared data are from the MIRLIN
observations. The 2.2$\mu$m and 8.4 GHz data are from S99 and C91
respectively. The hatched circles show the FWHM of the PSF for the 
observations.

Figure 1d. Images of Markarian 231. The format is 12.5$\mu$m image -
left, 12.5$\mu$m PSF star - center, 8.4 GHz image - right. The source 
locations and 
contours are set as in Figure 1a.The  mid-infrared data are from the
LWS observations. The  8.4 GHz data are from C91. The hatched circle in
the 8.4 GHz panel shows the FWHM of the PSF for this observation.

Figure 1e. Images of Markarian 273. The format, source locations  and
 contour levels are the same as in Figure 1c. The  mid-infrared data
 are from the MIRLIN observations. The 2.2$\mu$m and 8.4 GHz data are
 from S99 and C91 respectively.  The $+$ symbols are at the same
 coordinates in each panel, and are centered on the peaks of the
 2.2$\mu$m image. There is excellent agreement between the locations
 of the 2.2$\mu$m and 12.5$\mu$m peaks. The discrepancy between the
 positions of the infrared peaks and the radio peak is real (see
 text).  The hatched circles show the FWHM of the PSF for the
 observations.

Figure 1f. Images of Arp 220. The format, source locations and 
contour levels are the
same as in Figure 1c. The  mid-infrared data are from the MIRLIN
observations and are from Soifer et al. (1999). The 2.2$\mu$m and 
8.4 GHz data are from S99 and C91
respectively. The hatched circle in the 12.5$\mu$m panel shows the FWHM 
of the PSF for the observation.

Figure 1g. Images of IRAS~17208-0014.  The left image  is 2.2$\mu$m,
the  right image is 12.5$\mu$m. The source locations and contour 
levels are the same as in
Figure 1a. The  mid-infrared data are from the MIRLIN
observations. The 2.2$\mu$m  data are from S99. The hatched circles 
show the FWHM of the PSF for the observations.

\figcaption{} The curves of growth for the objects (solid) and PSF's
(dashed) for the five ULIRGS which are single nuclei. In all five the
PSF's have been normalized so the enclosed flux of the PSF within a
radius of 2.5$''$ equals that of the object. For the objects, the sky
level was set to be the value in an annulus between 2.5 and 3$''$.  In
the plots for  IRAS~08572+3915 and UGC 5101 two normalizations  are
used for the PSF; in the lower curve the PSF  is set to match the
central profile of the galaxy, while the  upper curve matches the
enclosed flux from the PSF and the galaxy at the outer radius of the
measurement beams, as for the other galaxies.  The data of IRAS0857+39
and Markarian 231 were obtained with LWS as outlined in the text; the
other three data sets are from MIRLIN observations.

\figcaption{} The spectral energy distributions of the seven ULIRGS
studied are shown as a function of the rest wavelength from $\sim$1
$\mu$m to 1 mm.  Filled circles represent the MIRLIN data, while open
circles have been taken from Carico et al. (1988) (near infrared data,
5$''$ diameter beam) and Soifer et al. (1989) (IRAS data
$\sim$1$'$ diameter beam). The MIRLIN photometry has been calculated
for a 4$''$ diameter beam centered on the object or between the two
nuclei if multiple nuclei are present. As discussed in the text, the
photometry is based on IRAS standards.

\figcaption{} Montage of images of Markarian 273. The images at
wavelengths from 7.9$\mu$m through 17.9$\mu$m obtained with
MIRLIN. The image at 3.4$\mu$m was obtained on the 200-inch Hale
Telescope using the Cassegrain infrared camera.  The images are
centered so that the northern peak is at the location 0,0 in each
panel.  The locations of the boxes in the 12.5$\mu$m image show where
the photometry of the peaks was obtained in all the images.

\figcaption{} Spectral energy distributions of the separate peaks in
Markarian 273 as determined from the images presented in Figure 4.
The total flux (also presented in Figure 3b) was measured in a 4$''$ diameter 
beam centered between
the peaks  and was distributed between the two
sources by doing photometry in non-overlapping boxes (displayed in the 
12.5$\mu$m panel of Figure 4) centered on the
two peaks.  The uncertainties shown  are statistical. The systematic
uncertainty depends on the wavelength, but is probably in the range 15
- 20 \%.

\clearpage
\normalsize
\newpage

\normalsize

%\begin{references}

\clearpage
\begin{table}

\centerline {Table 1}

%\centerline 
\caption{ Basic Properties of ULIRGS Observed }

\smallskip
\begin{tabular}{l c c c c }   
\tableline\tableline
Name & z & log L & Spectrum &  linear scale \\
  &  &   L$_{bol}$[L$_{\odot}$] &  & Kpc/$''$ \\
\tableline

IRAS05189-2524 &    0.043   & 12.08   & Sey 1 &  830 \\

IRAS08572+3915 &    0.058   & 12.10   & LINER &  1200 \\

UGC 5101 &    0.040   & 12.00   & LINER &   820 \\

Markarian 231 &    0.040   & 12.52   & Sey 1 &   800 \\

Markarian 273 &    0.037   & 12.13   & Sey  &   800 \\

Arp 220 &    0.018   & 12.17   & LINER & 360  \\

IRAS17208-0014 &    0.040   & 12.46   & HII &  800 \\

\tableline

\end{tabular}

\end{table}

\clearpage
%\landscape
\begin{table}

\centerline {Table 2}
%\caption {Table 2}

%\centerline 
\caption{ Log of Keck II mid-infrared Observations of ULIRGS}

\smallskip
\begin{tabular}{l c c c c c c c c }   
\tableline\tableline
Object & 7.9$\mu$m & 8.8$\mu$m & 9.7$\mu$m & 10.3$\mu$m & 11.7$\mu$m & 
12.5$\mu$m & 17.9$\mu$m & 24.5$\mu$m  \\
 
\tableline

IRAS05189& &  &  &  &  & $\surd$ &  & $\surd$ \\

IRAS08572&  $\surd$ & $\surd$ & $\surd$ &  &  & $\surd$\tablenotemark{*} & 
$\surd$ &   \\

UGC 5101 &   &  &  &  &  & $\surd$\tablenotemark{*} & $\surd$ &  \\

Mrk 231 &  $\surd$ & $\surd$ & $\surd$ & $\surd$ &$\surd$  & 
$\surd$\tablenotemark{*} & $\surd$ &   \\

Mrk 273& $\surd$  & $\surd$ & $\surd$ & $\surd$ & $\surd$ & $\surd$ & 
$\surd$ &  \\

Arp 220 &  $\surd$  & $\surd$ & $\surd$ & $\surd$ & $\surd$ & $\surd$ & 
$\surd$ & $\surd$ \\

IRAS17208& $\surd$  &  &  &  &  & $\surd$ &  &  \\

\tableline
\tablenotetext{*} {size determination made with the Long Wavelength 
Spectrograph in March 99}
\end{tabular}

\end{table}

\clearpage
%\portrait
\begin{table}
%\small
\scriptsize
\centerline {Table 3}

%\centerline 
\caption{ MIRLIN and IRAS Flux Densities for  ULIRGS}

\smallskip
\begin{tabular}{l r r r r r r r r r r r }   
\tableline\tableline
Object & 7.9$\mu$m & 8.8$\mu$m & 9.7$\mu$m & 10.3$\mu$m & 11.7$\mu$m & 
12.5$\mu$m &
12.5$\mu$m\tablenotemark{*} & 17.9$\mu$m & 24.5$\mu$m & 
25$\mu$m\tablenotemark{*}& 60$\mu$m\tablenotemark{*} \\
  & \multicolumn{8}{c}{ [mJy]} & \multicolumn{3}{c}{ [Jy]} \\
\tableline

IRAS05189& &  &  &  &  & 720 & 740 &  & 3.2$\pm$0.2 & 3.5 & 13.9 \\

IRAS08572&  615$\pm$18 & 434$\pm$10 & 221$\pm$10 &  &  & 218$\pm$10 & 340 & 
552$\pm$46 &  & 1.8 & 7.5 \\

UGC 5101&   &  &  &  &  & 185$\pm$10 & 260  & 256$\pm$42 &  & 1.1 & 12.1  \\

Mrk 231&  1370$\pm$43 & 1488$\pm$18 & 1081$\pm$30 & 1075$\pm$21 &1355$\pm
$15  & 1836$\pm$20 & 1930 & 3249$\pm$49 &  & 8.8 & 33.6  \\

Mrk 273& 267$\pm$15  & 141$\pm$10 & 92$\pm$10 & 38$\pm$6 & 150$\pm$5 & 210
$\pm$10 & 230 & 442$\pm$31 &  & 2.3 & 23.7 \\

Arp 220&  521$\pm$15  & 155$\pm$6 & 124$\pm$14 & 59$\pm$10 & 175$\pm$8 & 404
$\pm$12 & 500 & 1170$\pm$50 & 9.8$\pm$0.2 & 8.0 & 104.0\\

IRAS17208& 315$\pm$25  &  &  &  &  & 202$\pm$13 & 200  &  &  & 1.7 & 34.1\\

\tableline
\tablenotetext{} {The MIRLIN photometry refers to a 4$''$ diameter beam. 
The uncertainties listed in the table are statistical only. Photometric 
uncertainties in the MIRLIN data are $\pm$5\% for $\lambda <$ 20$\mu$m, 
and $\pm$10\% for $\lambda > $ 20$\mu$m. The uncertainties in the IRAS 
data are 5-10\%.}
\tablenotetext{*} {Flux densities from IRAS Catalog and are not color 
corrected}
\end{tabular}

\end{table}

\clearpage
%\portrait

\begin{table}
\small
\centerline {Table 4}

\caption{ Properties of Compact Sources in ULIRGS}

\smallskip
\begin{tabular}{l c c c c c c c c  }   
\tableline\tableline
Object & \multicolumn{2}{c}{12.5$\mu$m} & \multicolumn{2}{c} {17.9$\mu$m} 
&   \multicolumn{2}{c} {24.5$\mu$m} & 8.4~Ghz & 2.2~$\mu$m\\

  &   \% flux\tablenotemark{+} &  size &    \% flux\tablenotemark{+} &  
size & \% flux\tablenotemark{+} &  size  & size & size \\

  &   in core &$''$ & in core &$''$ & in core & $''$ &$''$ &$''$ \\
\tableline

IRAS05189&100 &$ <$ 0.20 & -- &--   & 100 &  $<$0.26 & $0.20\times0.17$ & 
$<$0.2\\

IRAS08572& $\sim$80 & $<$0.22\tablenotemark{*} & $\sim$100 &$<$0.25 &--  
& -- & $0.09\times0.07$ & $<$0.2\\

UGC 5101& 50 & $<$0.25\tablenotemark{*} & 60 &$<$0.25 &--   &--  & 
$0.14\times0.11$ &$<$0.2\\

Mrk 231& 100 & $<$0.13\tablenotemark{*} & 85 &$<$0.25 &--   & --&
 $\le0.07\times0.06$ & $<$0.2\\ 

Mrk 273N & 67  & $<$0.3 & -- &-- &-- &-- &$0.32\times 0.18$& $\sim$0.3\\

Mrk 273SW& 50  & $<$0.3 & -- &-- &-- &-- & $0.18\times0.11$ & $<$0.2\\

Arp 220 E&-- &-- &-- &-- &25 & -- & $0.32\times0.19$ & $<$0.2 \\
Arp 220 W&-- &-- &-- &-- &75 &$0.40$ & $0.21 \times 0.14$ & 0.4 \\

IRAS17208& 100  & 2.0 & -- & -- &--  &-- & $0.26 \times 0.32$
\tablenotemark{x} & 2  \\

\tableline
\tablenotetext{+} {percentage of flux measured in 4$''$ diameter beam}
\tablenotetext{*} {size measurement from LWS observations}
\tablenotetext{x} {size measurement at 1.6 GHz from  Martin et al. 1989}
\end{tabular}

\end{table}

\clearpage
\begin{table}
\small
\centerline {Table 5}

\caption{ Infrared Sizes of ULRIGs }

\smallskip
\begin{tabular}{l c c r r }   
\tableline\tableline
Object &\multicolumn{2}{c} {size of mid-ir core} & \multicolumn{2}{c} 
{L/A[L$_{\odot}$Kpc$^{-2}$]}  \\

       & $''$ & pc &  max  & min  \\

\tableline

IRAS05189& $<$0.2& $<$166 & $5.6\times 10^{13}$ & $2.1\times 10^{13}$\\

IRAS08572& $<$0.22& $<$260 & $>2.8\times 10^{13}$ & $>6.0\times 10^{12}$ \\

UGC 5101& $<$0.25& $<$205 & $>1.1\times 10^{13}$ & $>1.0\times 10^{12}$\\

Mrk 231&   $ <$0.13& $<$104 & $3.9\times 10^{14}$ & $1.0\times 10^{14}$\\

Mrk 273N&   $<$0.3 & $<$240  & $>2.2\times 10^{13}$ & $>1.0\times 10^{12}$ \\

Arp 220&   0.40 & 140 & $6.0\times 10^{13}$ & $1.2\times 10^{13}$\\

IRAS17208& 2.0& 1600 & $1.2\times 10^{12}$ & $3.0\times 10^{10}$\\

\tableline

\end{tabular}

\end{table}

\clearpage
\begin{table}
\small
\centerline {Table 6}

\caption{ Summary of Properties of  ULIRGS }

\smallskip
\begin{tabular}{l  c c c c c c  }   
\tableline\tableline
Object & {Compact core?} & \multicolumn{4}{c} {  Spectral Class}   \\

       &  MIR/NIR/Radio & Optical & Near IR & ISO & ASCA \\

\tableline

IRAS05189& Y/Y/N & Sey2 & Sey & -- & AGN\\

IRAS08572& Y/Y/N & LINER & HII& --& SB \\

UGC 5101& Y/Y/Y & LINER & --& SB & SB\\

Mrk 231&   Y/Y/Y & Sey1 & Sey1 & AGN & SB?\\

Mrk 273&   Y/Y/Y & Sey2 &  Sey? & AGN & AGN\\

Arp 220&   Y/N/Y & LINER & --& SB & SB?\\

IRAS17208&  N/N/? & HII & HII & SB & --\\

\tableline

\end{tabular}

\end{table}

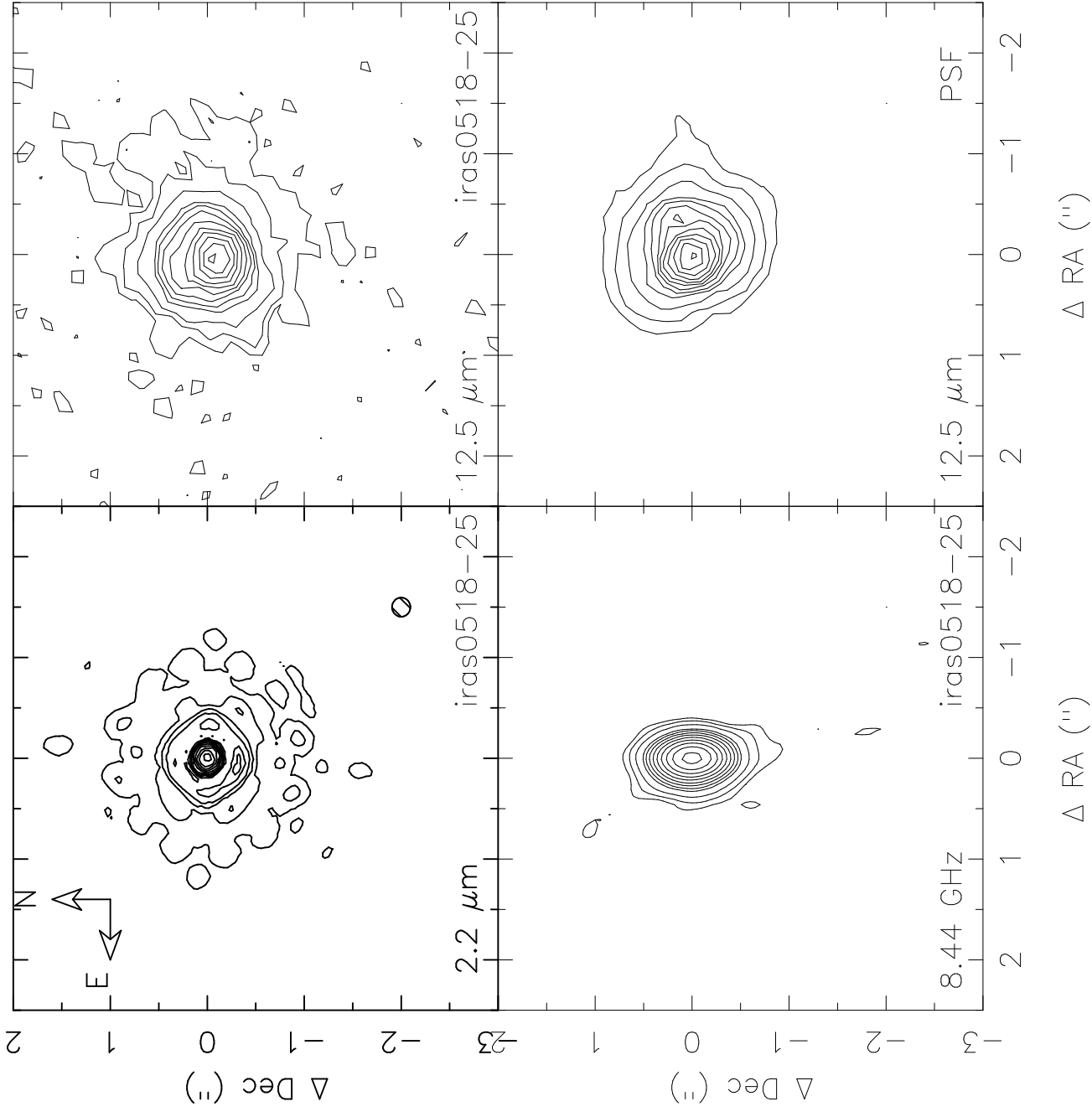
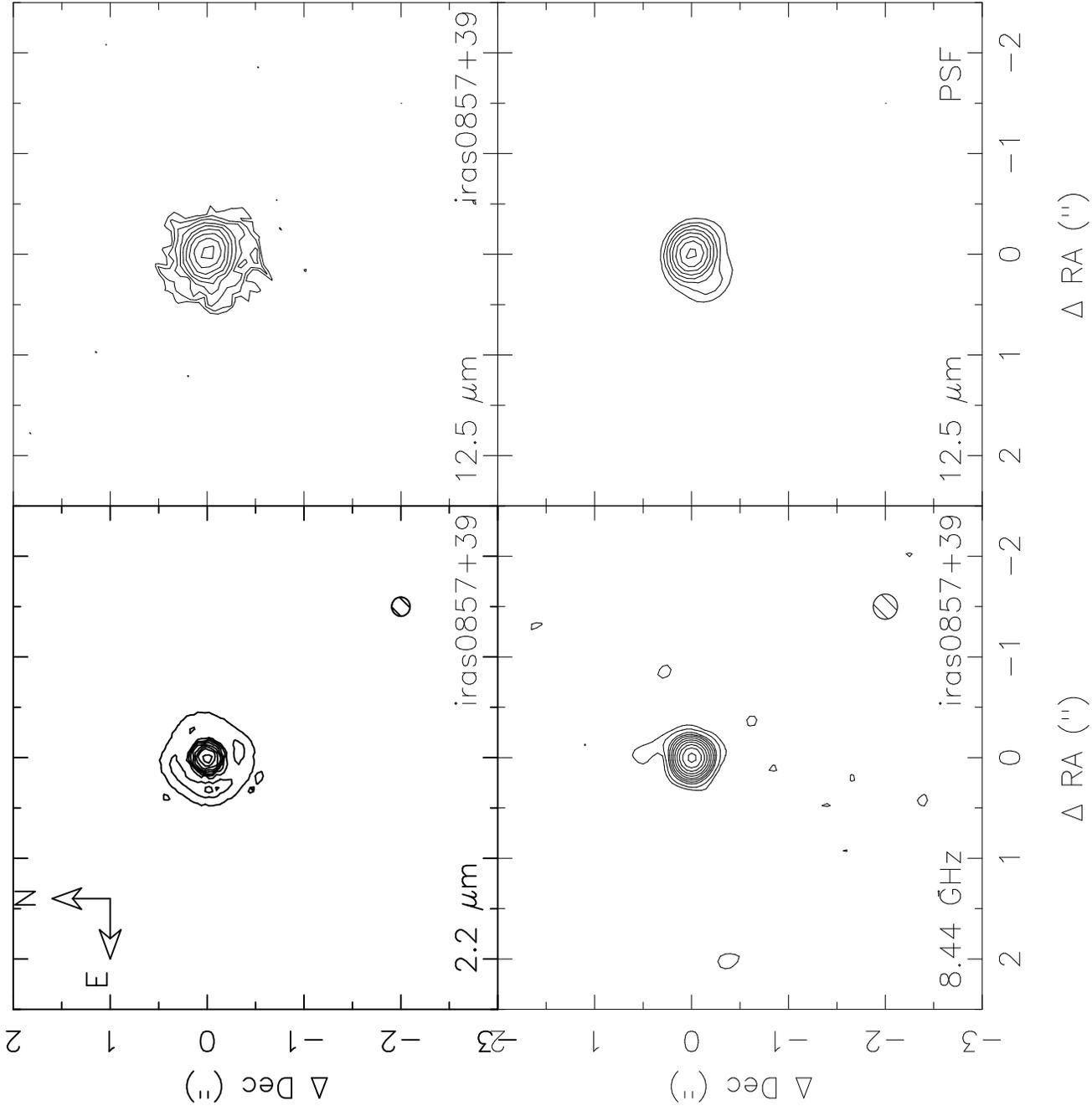
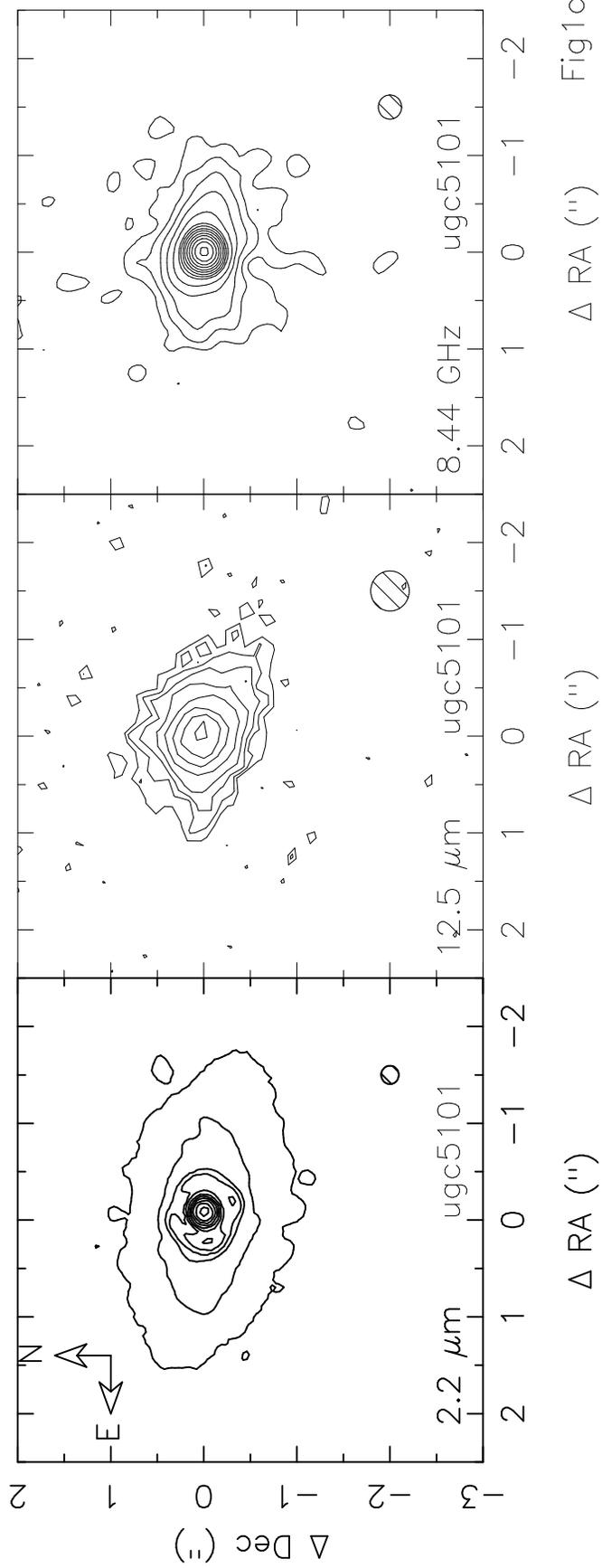
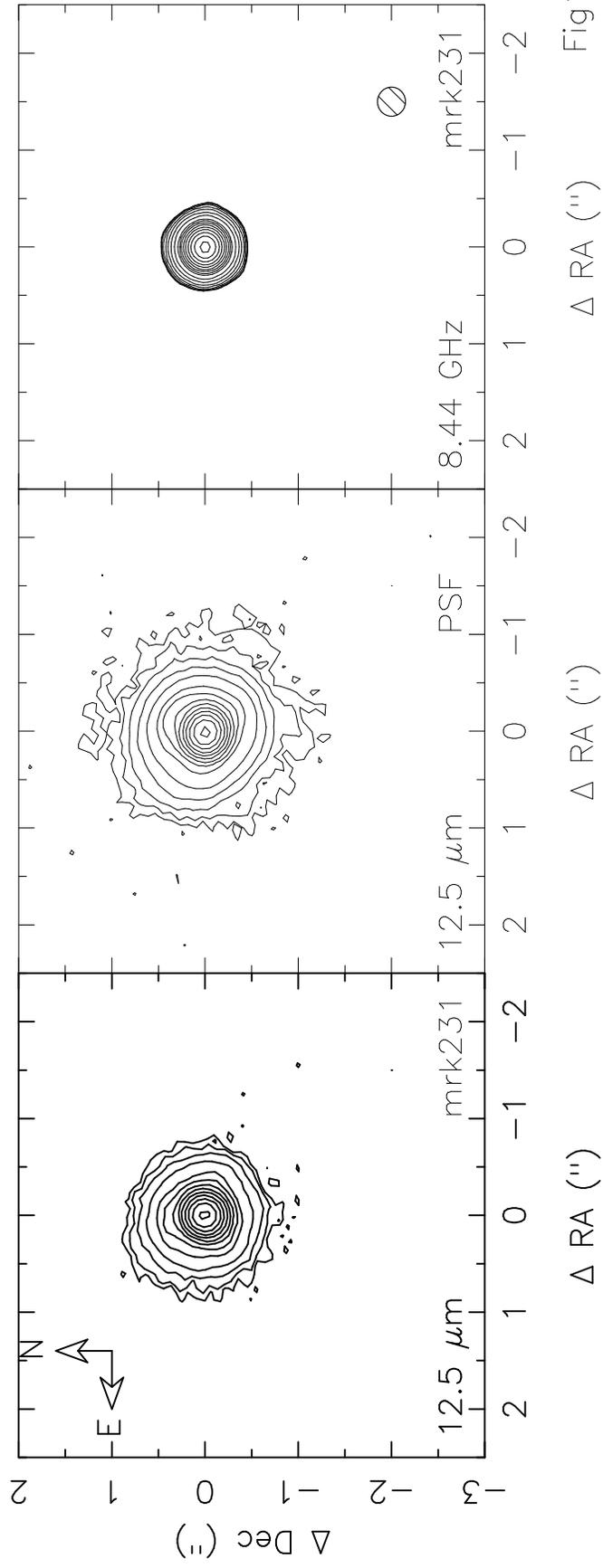
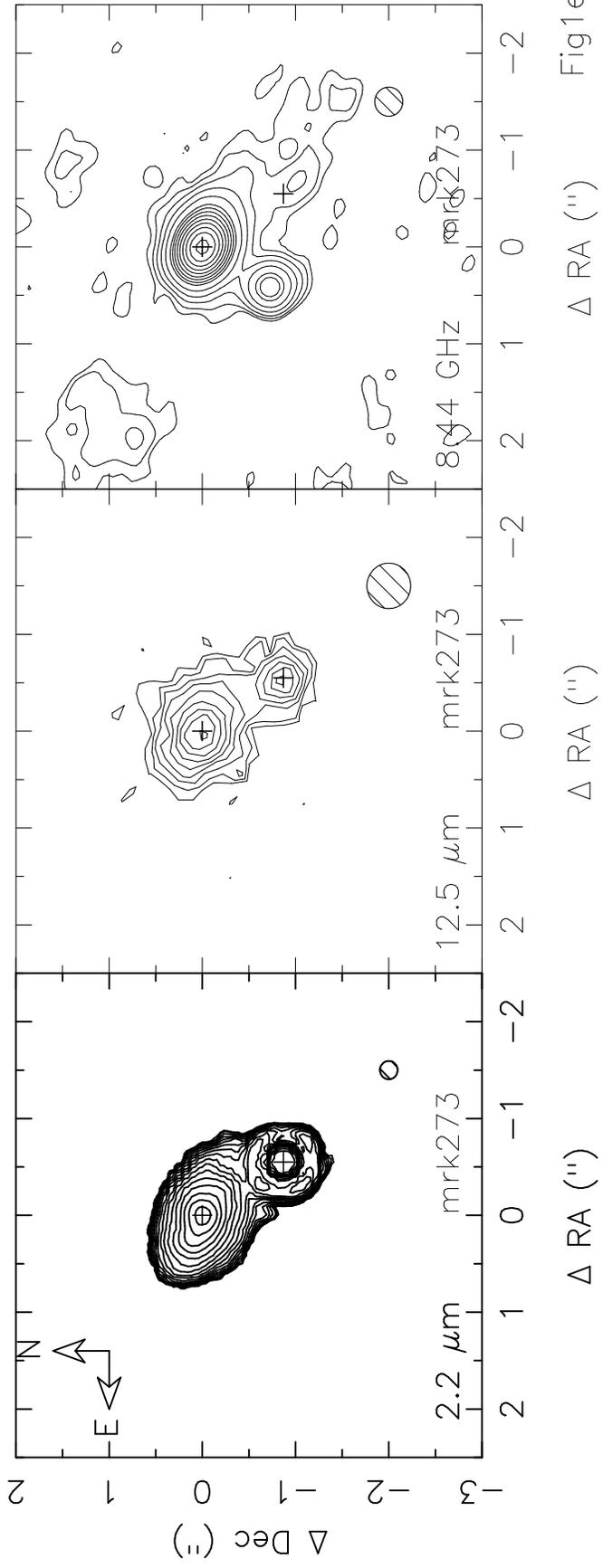
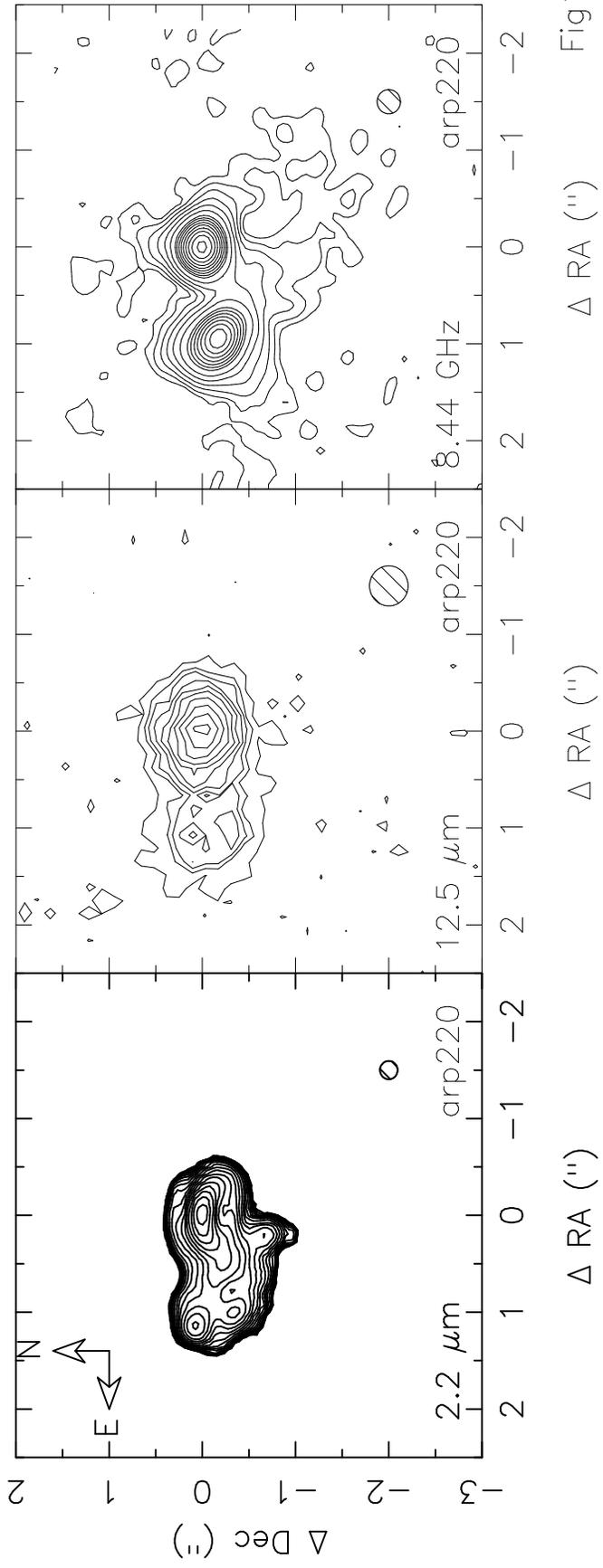
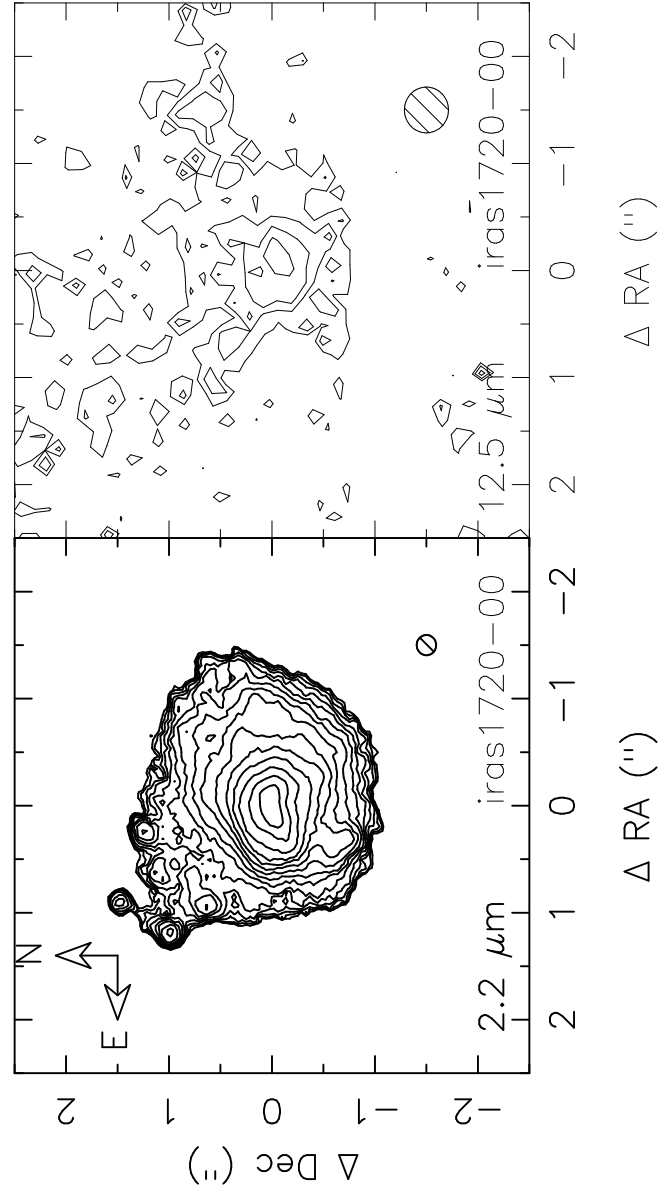
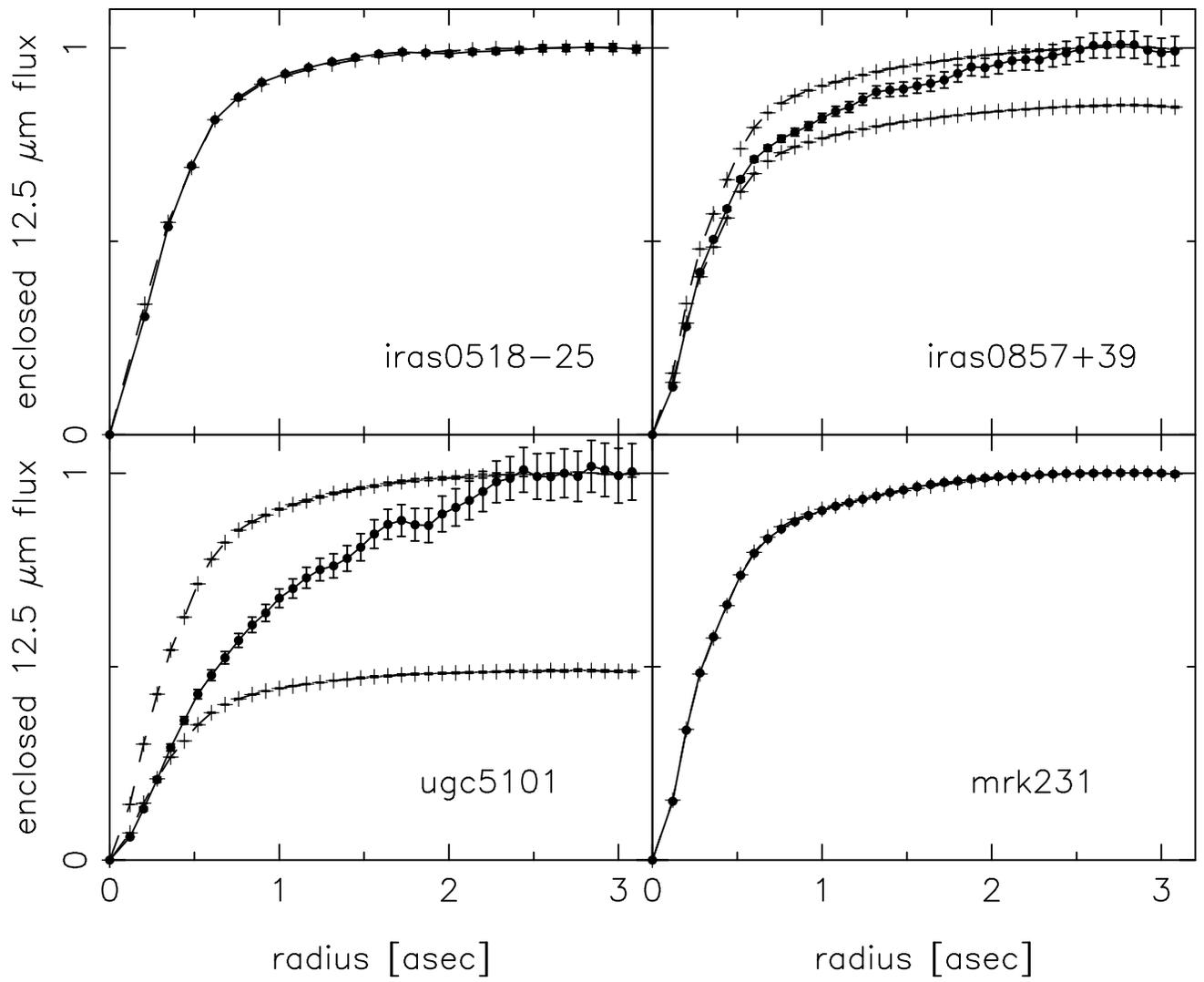
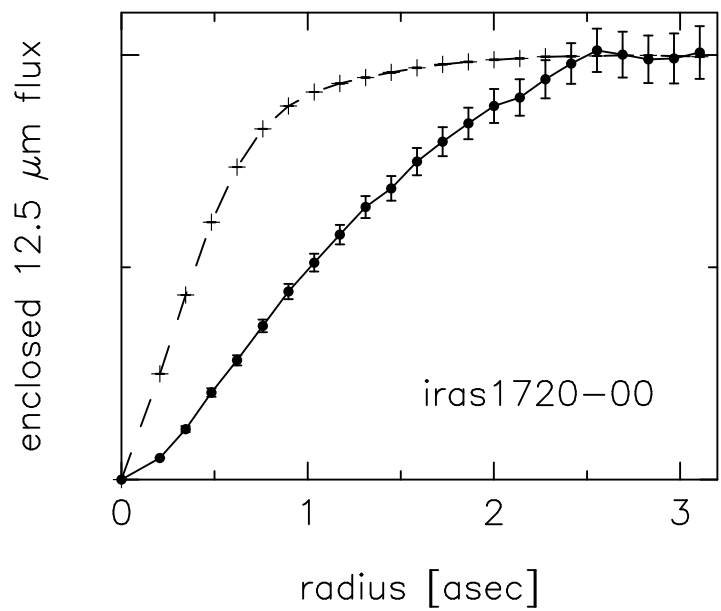
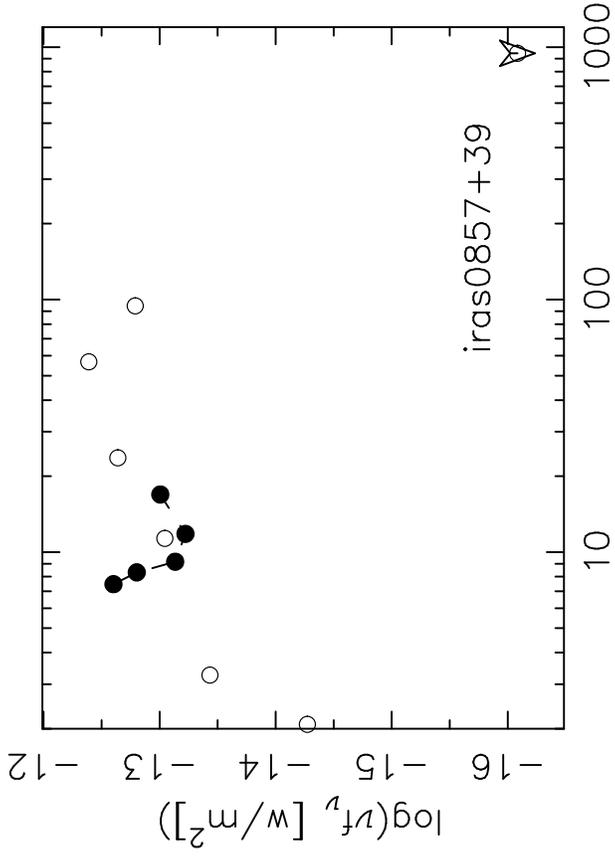
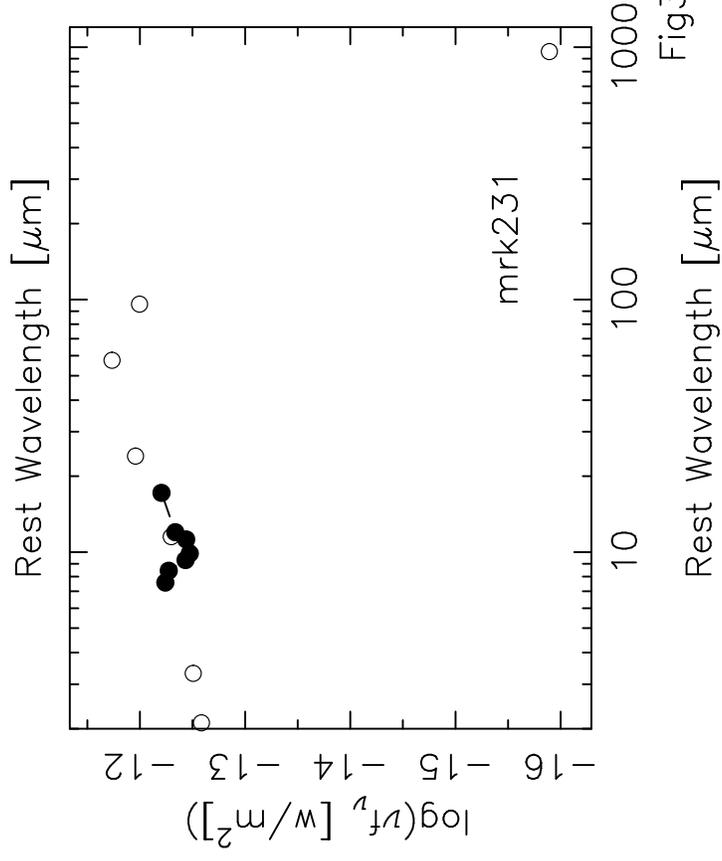
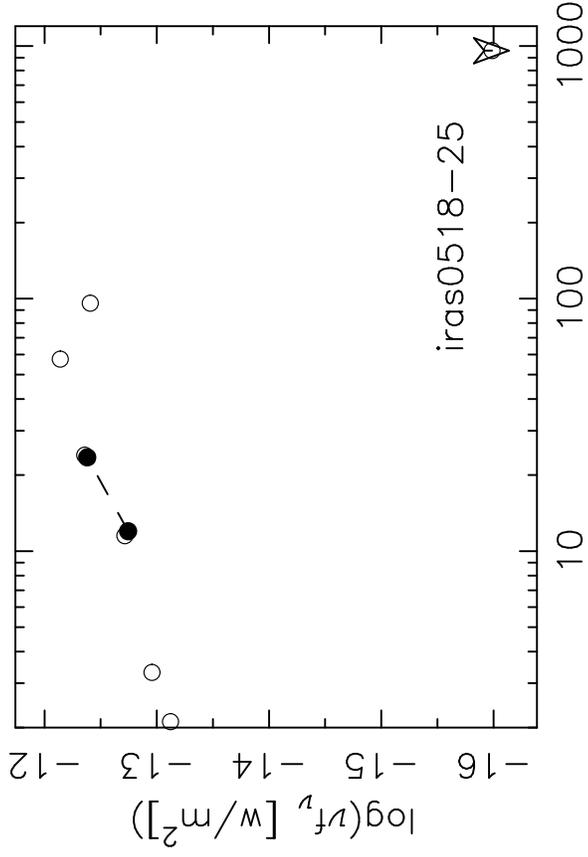
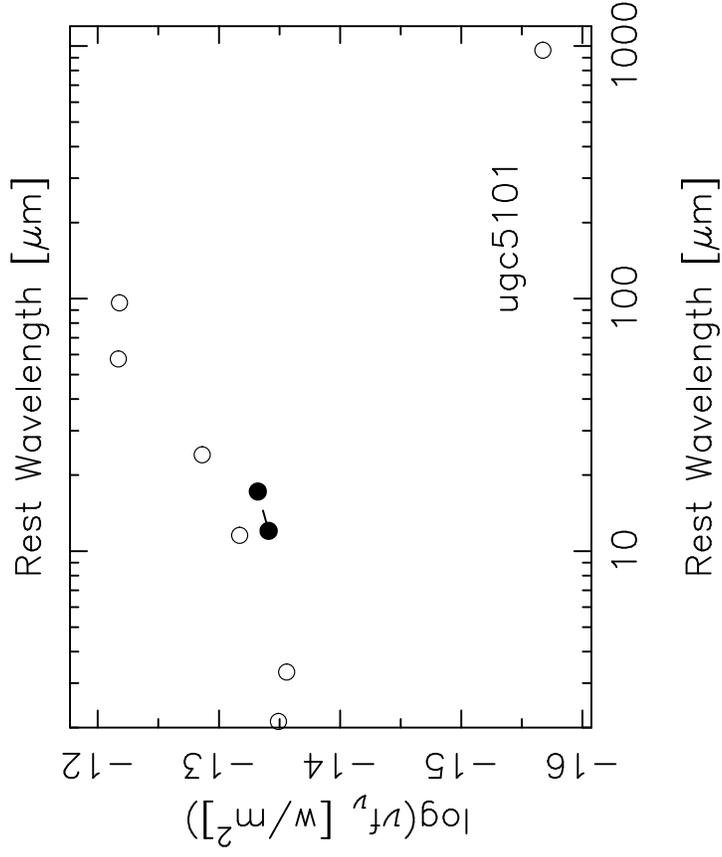
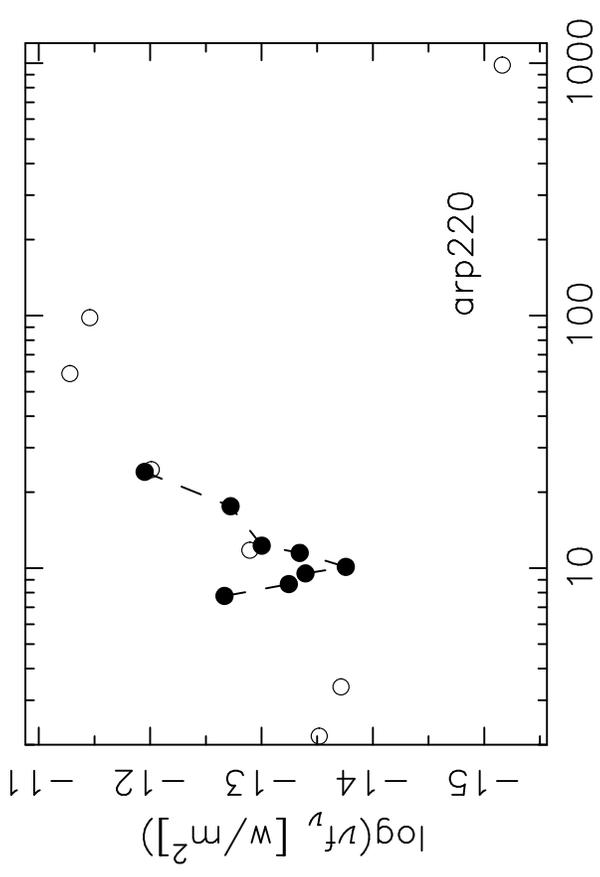
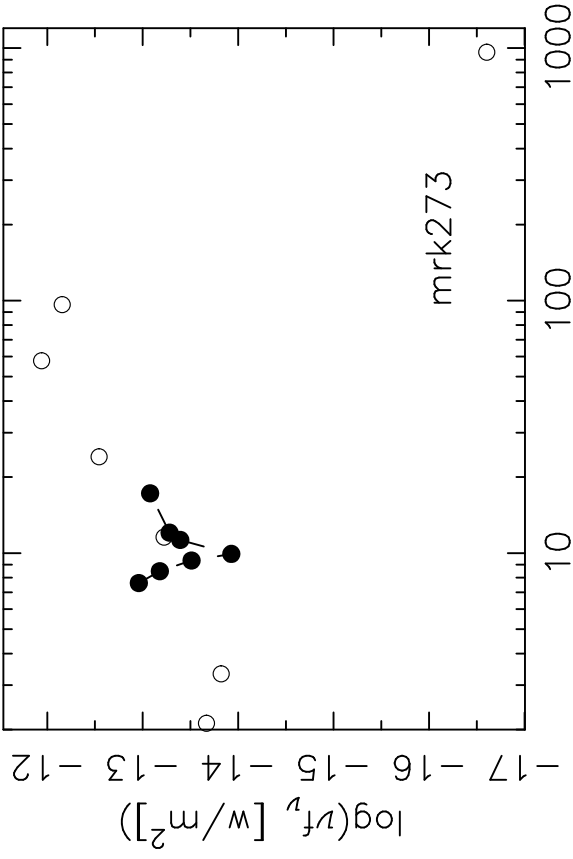
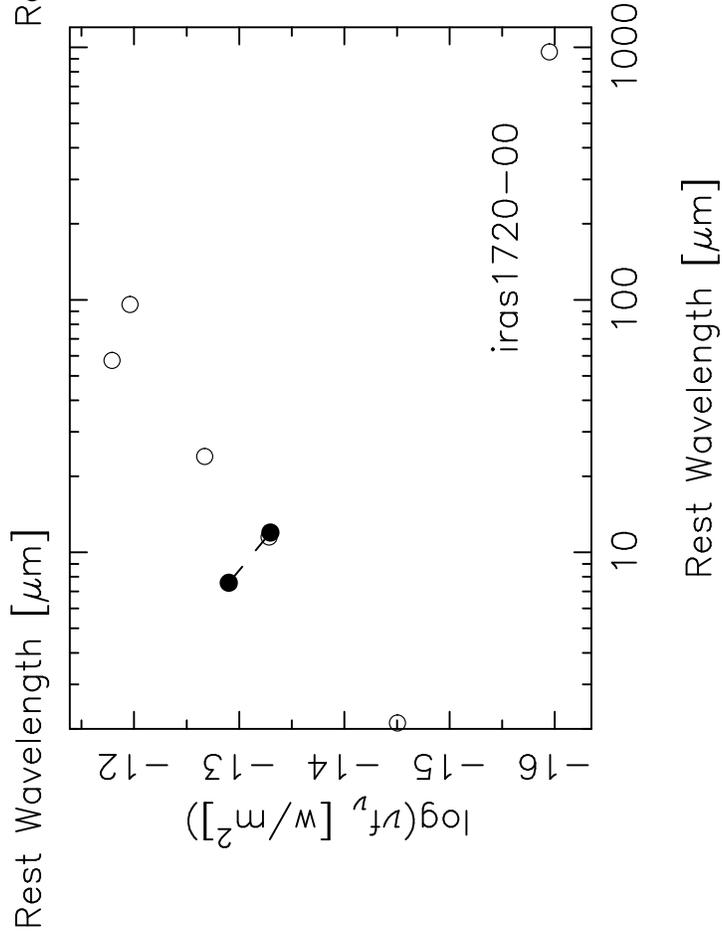
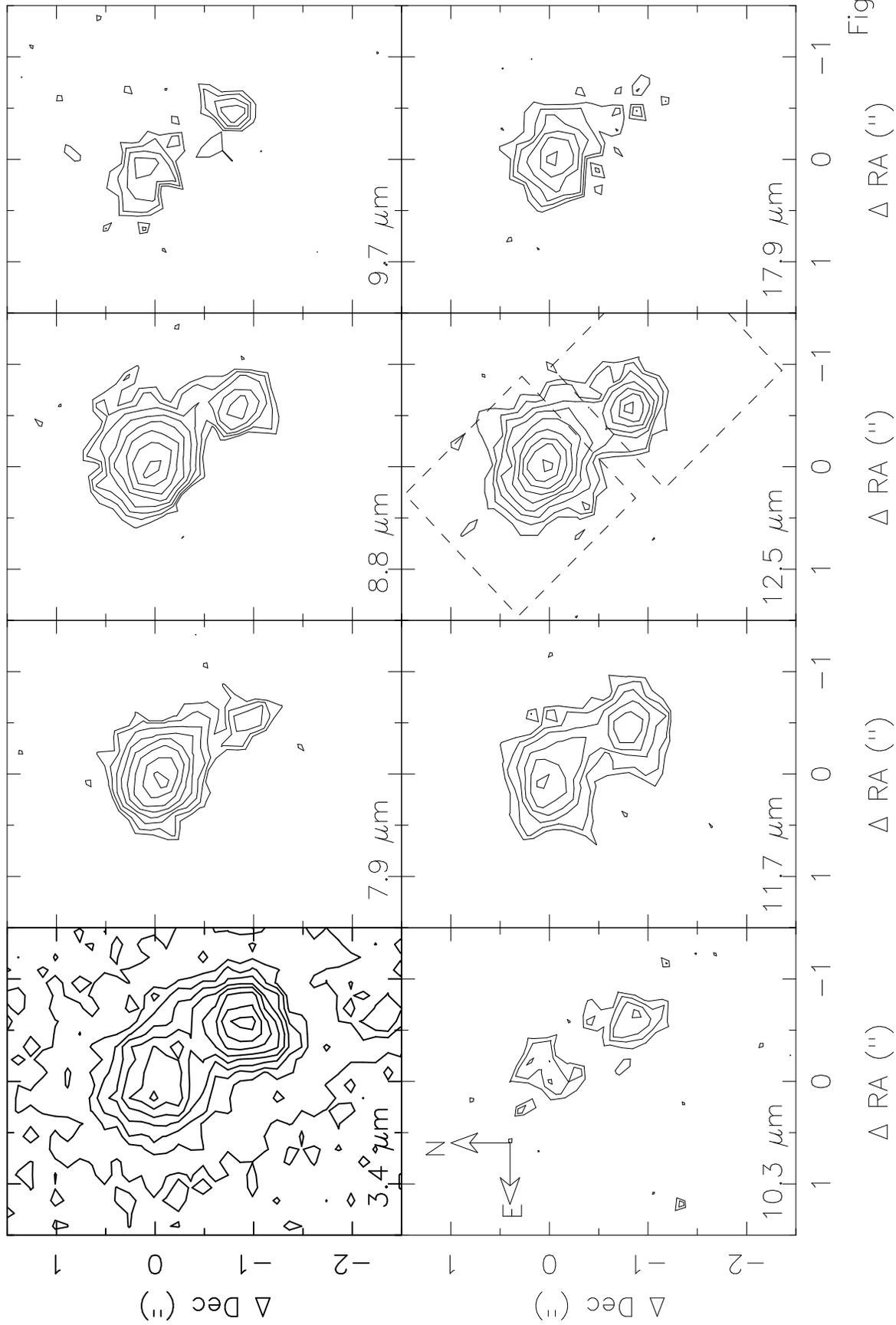
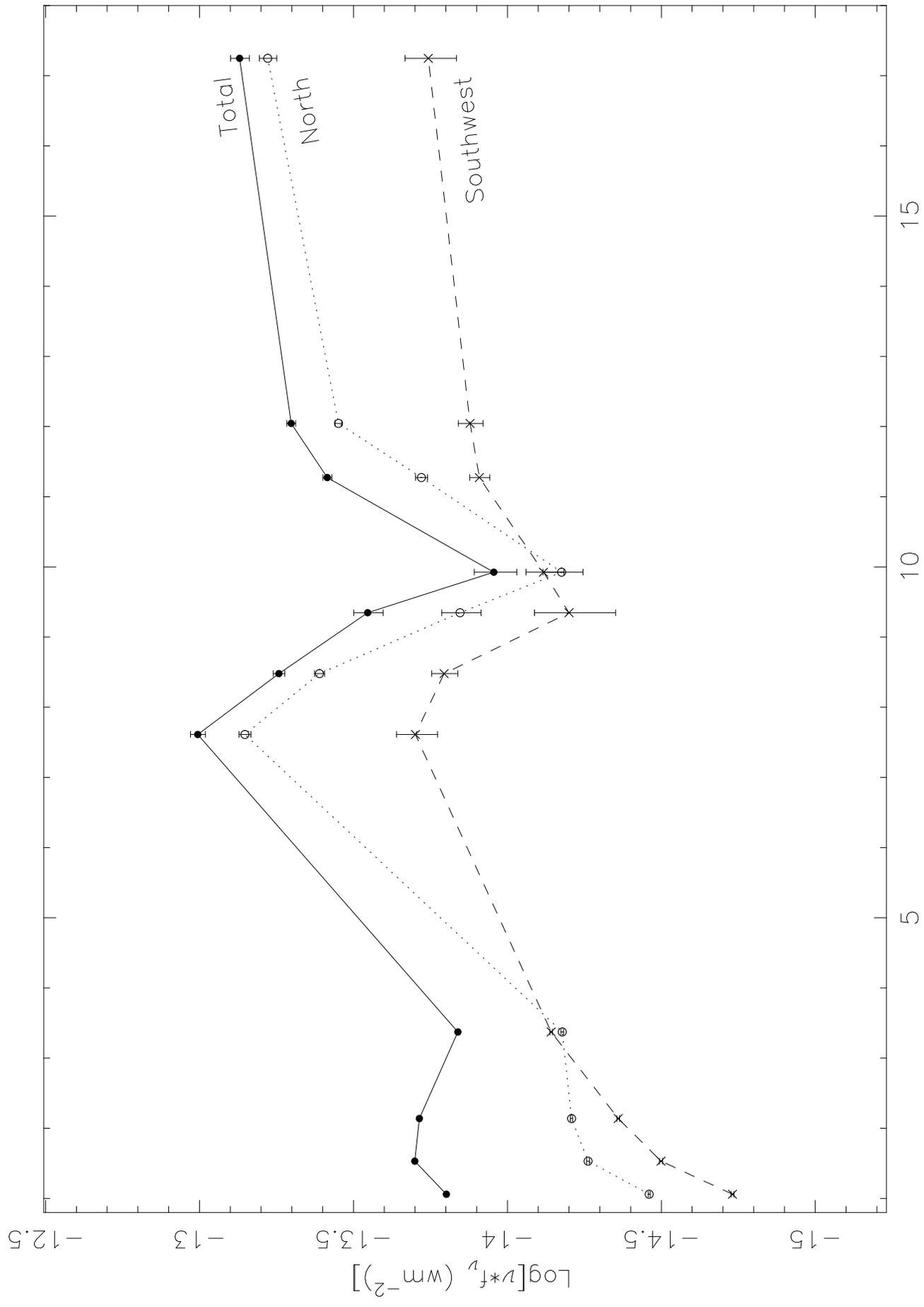
\end{document}